\documentclass[twocolumn]{aastex63}

\usepackage{amsmath}
\usepackage{autobreak}

\shorttitle{NGC\,4214 X-1}
\shortauthors{Lin, Soria, and Swartz}

\graphicspath{{./}{figures/}}

\begin{document}

\title{On the Short-Period Eclipsing High-Mass X-ray Binary in NGC\,4214}

\author[0000-0001-9576-1870]{Zikun Lin}
\affiliation{Key Laboratory of Optical Astronomy, National Astronomical Observatories, Chinese Academy of Sciences,\\ Beijing 100101, China}
\affiliation{School of Astronomy and Space Science, University of Chinese Academy of Sciences, Beijing 100049, China}

\author[0000-0002-4622-796X]{Roberto Soria}
\affiliation{School of Astronomy and Space Science, University of Chinese Academy of Sciences, Beijing 100049, China}
\affiliation{INAF -- Osservatorio Astrofisico di Torino, Strada Osservatorio 20, I-10025 Pino Torinese, Italy}
\affiliation{Sydney Institute for Astronomy, School of Physics A28, The University of Sydney, Sydney, NSW 2006, Australia}
\author[0000-0002-2954-4461]{Douglas A. Swartz}
\affiliation{Science and Technology Institute, Universities Space Research Association, Huntsville, AL 35805, USA}

\begin{abstract}
We present the results of our study of the luminous ($L_{\rm X} \approx 10^{39}$ erg s$^{-1}$) X-ray binary CXOU\,J121538.2$+$361921 in NGC\,4214, the high mass X-ray binary with the shortest known orbital period. Using {\it Chandra} data, we confirm the $\approx$13,000 s (3.6 hr) eclipse period, and an eclipse duration of $\approx$2000 s. From this, we estimate a mass ratio $M_2/M_1 \gtrsim 3$ and a stellar density $\rho \approx 6$ g cm$^{-3}$, which implies that the donor must be a Wolf-Rayet or a stripped Helium star. The eclipse egress is consistently much slower than the ingress. This can be explained by denser gas located either in front of the compact object (as expected for a bow shock) or trailing the donor star (as expected for a shadow wind, launched from the shaded side of the donor). There is no change in X-ray spectral shape with changing flux during the egress, which suggests either variable partial covering of the X-ray source by opaque clumps or, more likely, a grey opacity dominated by electron scattering in a highly ionized medium. We identify the optical counterpart from {\it Hubble} images. Photometry blueward of $\sim$5500~\AA\ indicates a bright ($M_{B} \approx -3.6 \pm 0.3$~mag, for a range of plausible extinctions), hot ($T \approx 90,000 \pm 30,000$ K) emitter, consistent with the Wolf-Rayet scenario. There is also a bright ($M_{I} \approx -5.2$~mag), cool ($T \approx 2700 \pm 300$ K) component consistent with an irradiated circumbinary disk or with a chance projection of an unrelated asymptotic giant branch star along the same line of sight. 
 
\end{abstract}

\section{Introduction} \label{sec:intro}

X-ray binaries provide a crucial empirical constraint for population synthesis models and for predictions of the merger rate in double compact binaries in the nearby universe. To this aim, it is important to identify X-ray binaries with a massive donor star (initial mass $\gtrsim 8 M_{\odot}$) and a short binary period ($\lesssim 2$ d), that is systems that can not only evolve into double compact binaries, but also merge via gravitational wave decay within a Hubble time (see \citealt{mandel22} for a recent review).

Satisfying those two conditions simultaneously is of course difficult, considering the large radii of main sequence OB stars, Be stars, and supergiants, which imply typical periods of several days or weeks in most high mass X-ray binaries (HMXBs) \citep{kretschmar19,vandenheuvel19,walter15}. A rare exception is represented by massive systems that have gone through a common envelope stage, in which the compact object (either a black hole (BH) or a neutron star (NS)) has been temporarily engulfed in the envelope of the expanding donor star, after the donor has overfilled its Roche lobe \citep{belczynski16,vandenheuvel17}. In this phase, the binary period can decrease by orders of magnitude, as the compact object and the core of the donor spiral towards each other. If the common envelope is blown away before a merger, the resulting system is likely to be an X-ray binary with a very short period and a hot donor star stripped of its hydrogen envelope. Such massive Helium-burning stars can loosely be classified as Wolf-Rayet (WR) stars, and the associated X-ray system is called a WR X-ray binary. Some authors \citep[{\it e.g.},][]{sander20a,sander20b,gotberg18} prefer to distinguish between classical WR stars, with masses above $\approx$20 $M_{\odot}$, which shed their hydrogen envelop through radiation-driven, optically thick wind, and lower-mass stripped He stars, with optically thin winds,  which lose their envelope through binary evolution. 
A common envelope phase is not the only way to produce  short-period HMXBs and double compact binaries: steady Roche-lobe overflow \citep{vandenheuvel17} and chemically homogeneous evolution of massive binary stars \citep{marchant17} have also been proposed. However, for HMXBs with binary periods as short as a few hours, a common envelope spiral-in phase remains the most likely channel.  

\begin{deluxetable*}{lccccc}\label{tab:sum_cand}
\tablecaption{Main properties of all the short-period candidate WR X-ray binaries known to-date \label{tab:WR_binary}}
\tablewidth{0pt}
\tablehead{
\colhead{Source Name}  & \colhead{Galaxy} & \colhead{Distance$^a$} & \colhead{Peak $L_{0.3-10}^b$} & \colhead{Period} & \colhead{References$^c$} \\[-5pt]
\colhead{}  & \colhead{} & \colhead{(Mpc)} & \colhead{(erg s$^{-1}$)} & \colhead{(hr)} & \colhead{}
}
\startdata
CXOU\,J121538.2$+$361921 & NGC\,4214 & 3.0 & $\approx$1$\times 10^{39}$ & 3.6 & 1, this work\\[2pt]
Cygnus X-3 & Milky Way & 0.0074 & $\approx$ a few $\times 10^{38}$ & 4.8 & 2,3,4,5,6,7\\[2pt]
CXOU\,J123030.3$+$413853 & NGC\,4490 & 6.5 & $\approx$1$\times 10^{39}$ & 6.4 & 8\\ [2pt]
CXOU\,J141312.2$-$652013 & Circinus & 4.2  &  $\approx$3$\times 10^{40}$ & 7.2 & 9,10,11\\ [2pt]
CXOU\,J004732.0$-$251722 & NGC\,253 & 3.5 & $\approx$1$\times 10^{38}$ & 14.5 & 12\\ [2pt]
CXOU\,J005510.0$-$374212  & NGC\,300 & 1.9 & $\approx$3$\times 10^{38}$ &   32.8 & 13,14,15,16,17 \\[2pt]
CXOU\,J002029.1$+$591651  & IC\,10 & 0.7 & $\approx$7$\times 10^{37}$ &   34.8 & 14,18,19,20,21,22 \\
\enddata
\tablecomments{
$^a$Redshift-independent distances from the NASA/IPAC Extragalactic Database. For each galaxy, we selected the median value of Cepheid and tip-of-the-red-giant-branch measurements, when available, otherwise Tully-Fisher distances.\\
$^b$De-absorbed 0.3--10 keV luminosity in the bright phase of the orbital cycle; values taken from the references listed in this Table, but rescaled to the distances adopted here when different.\\
$^c$References:
1: \cite{ghosh06};
2: \cite{lommen05};
3: \cite{hjalmarsdotter09};
4: \cite{koljonen10};
5: \cite{zdziarski12};
6: \cite{mccollough16};
7: \cite{veledina23}
8: \cite{esposito13};
9: \cite{weisskopf04};
10: \cite{esposito15};
11: \cite{qiu19b};
12: \cite{maccarone14};
13: \cite{carpano07};
14: \cite{barnard08};
15: \cite{crowther10};
16: \cite{binder11};
17: \cite{binder21};
18: \cite{prestwich07};
19: \cite{silverman08};
20: \cite{laycock15};
21: \cite{steiner16};
22: \cite{bhattacharya23}.
}
\end{deluxetable*}

An observational census of candidate WR/stripped X-ray binaries in nearby galaxies and individual modelling of their system parameters provide crucial constraints to such theoretical models of binary evolution and spiral-in processes \citep{vandenheuvel17}. The predicted number of such short-period X-ray binaries depends on the minimum mass ratio of donor star over compact object that triggers the formation of a common envelope instead of steady Roche-lobe overflow, and on the probability of survival of the common envelope phase for NSs and BHs of different masses. Observationally, very few candidates have been found so far in the Milky Way and nearby galaxies, which is an issue of theoretical concern \citep{vandenheuvel17,lommen05}.
The only Galactic system is the microquasar Cygnus X-3 \citep[{\it e.g.},][]{vandenheuvel73,lommen05,zdziarski12,zdziarski13,belczynski13,mccollough16,veledina23}, with a period of 4.8 hr. It has an X-ray luminosity of a few times $10^{38}$ erg s$^{-1}$, and shows repeated transitions between several X-ray/radio states \citep{szostek08,koljonen10}. The nature of its compact object is still unclear: either a massive NS at the highest end of its mass range, or, more likely, a low-mass BH \citep{zdziarski13}. Another six short-period candidate WR X-ray binaries have been found in external galaxies: their properties are summarized in Table \ref{tab:WR_binary} (adapted and updated from \citealt{esposito15,qiu19a,qiu19b}). Among them, the system with the shortest binary period is CXOU J121538.2$+$361921 in NGC\,4214 (henceforth, NGC\,4214 X-1 for simplicity). This is the target of this study.

\begin{figure}
  \centering
    \includegraphics[width=1\linewidth]{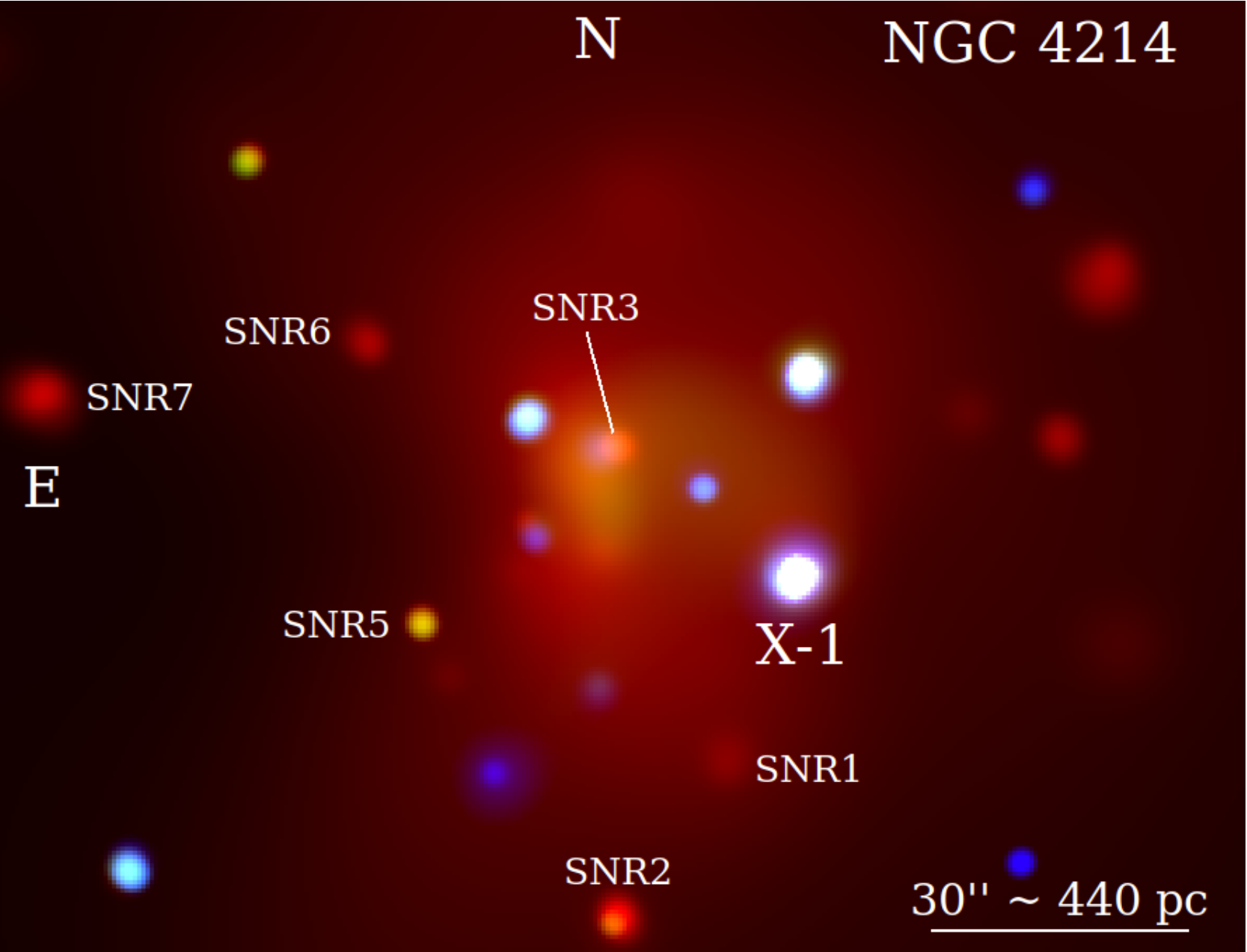}
    \includegraphics[width=1\linewidth]{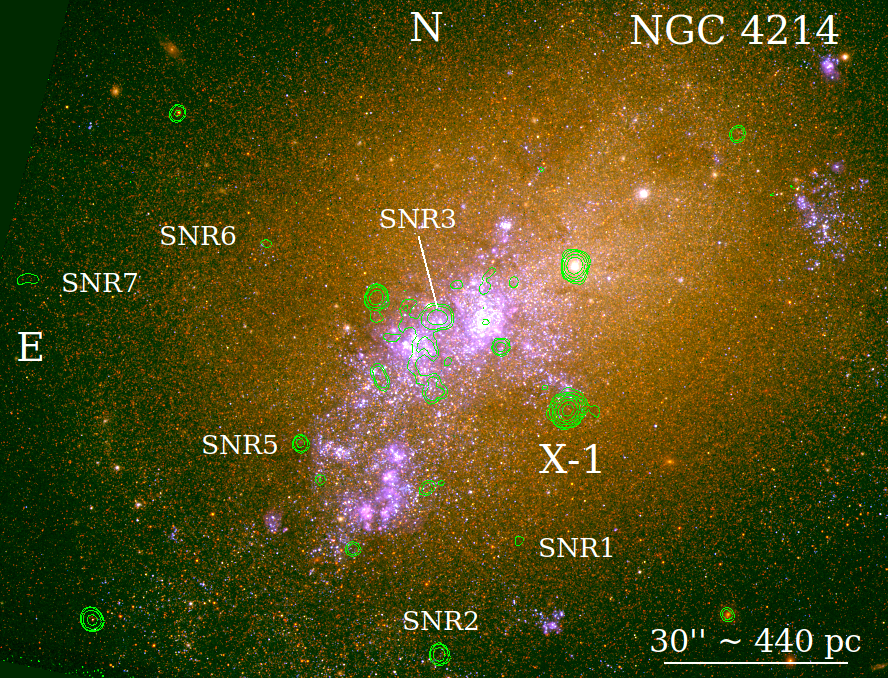}
    \caption{Top Panel: {\it Chandra}/ACIS true colour image of NGC\,4214, obtained from a stack of all four datasets and then adaptively smoothed. Red = 0.3--1.0 keV, green = 1.0--2.0 keV, blue = 2.0--7.0 keV. X-1 is the subject of this study. SNR1, SNR2, SNR3, SNR5, SNR6, and SNR7 are the X-ray counterparts of the optical SNRs identified by \cite{dopita10}; their estimated X-ray luminosities are listed in Table \ref{tab:SNR}. Bottom panel: {\it HST}/WFC3 UVIS true-color image on the same scale as the {\it Chandra} image. Red = F814W ($\approx${\it I}), green = F547M ($\approx${\it V}), blue = F336W ($\approx${\it U}). {\it Chandra} flux contours of the strongest X-ray sources are overplotted in green (0.3--7.0 keV band).}
    \label{fig1}
\end{figure}

\begin{figure}
  \centering
    \includegraphics[width=1\linewidth]{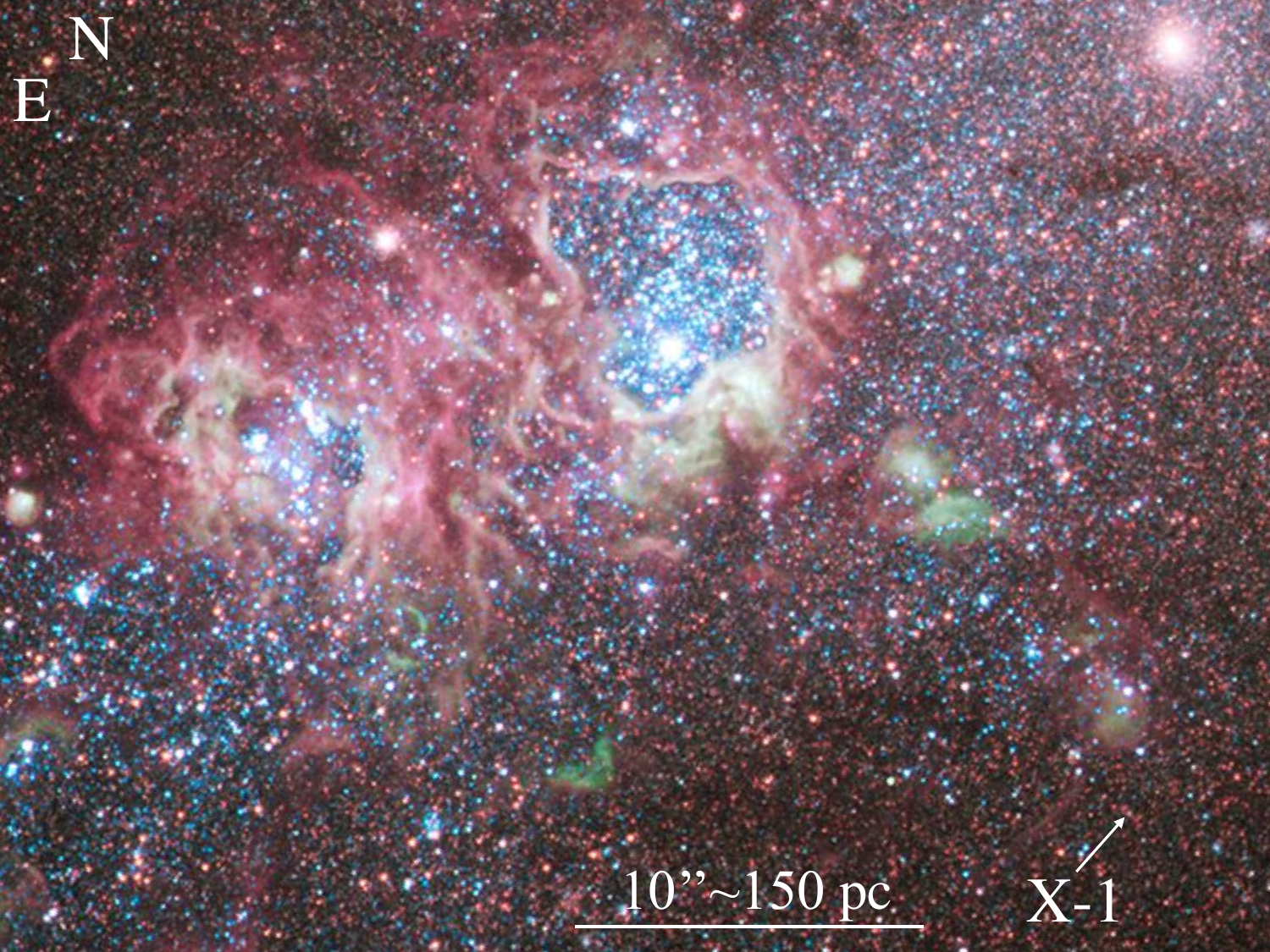}
    \caption{Details of the central region of NGC\,4214, adapted from an {\it HST}/WFC3 image processed and published by the Hubble Heritage Team (images from the GO-11360 program; PI: R.~O’Connell). The red color comes from a combination of the F657N and F814W filters; the green color is from F502N and F547M; the blue color is from F225W, F336W, F438W, and F487N. The optical counterpart of X-1 is marked by an arrow near the bottom right of the image. It is at the outskirts of the current starburst region. 
Credit: NASA, ESA, and the Hubble Heritage (STScI/AURA)-ESA/
Hubble Collaboration.}
    \label{fig1b}
\end{figure}

\begin{figure}
  \centering
    \includegraphics[width=1\linewidth]{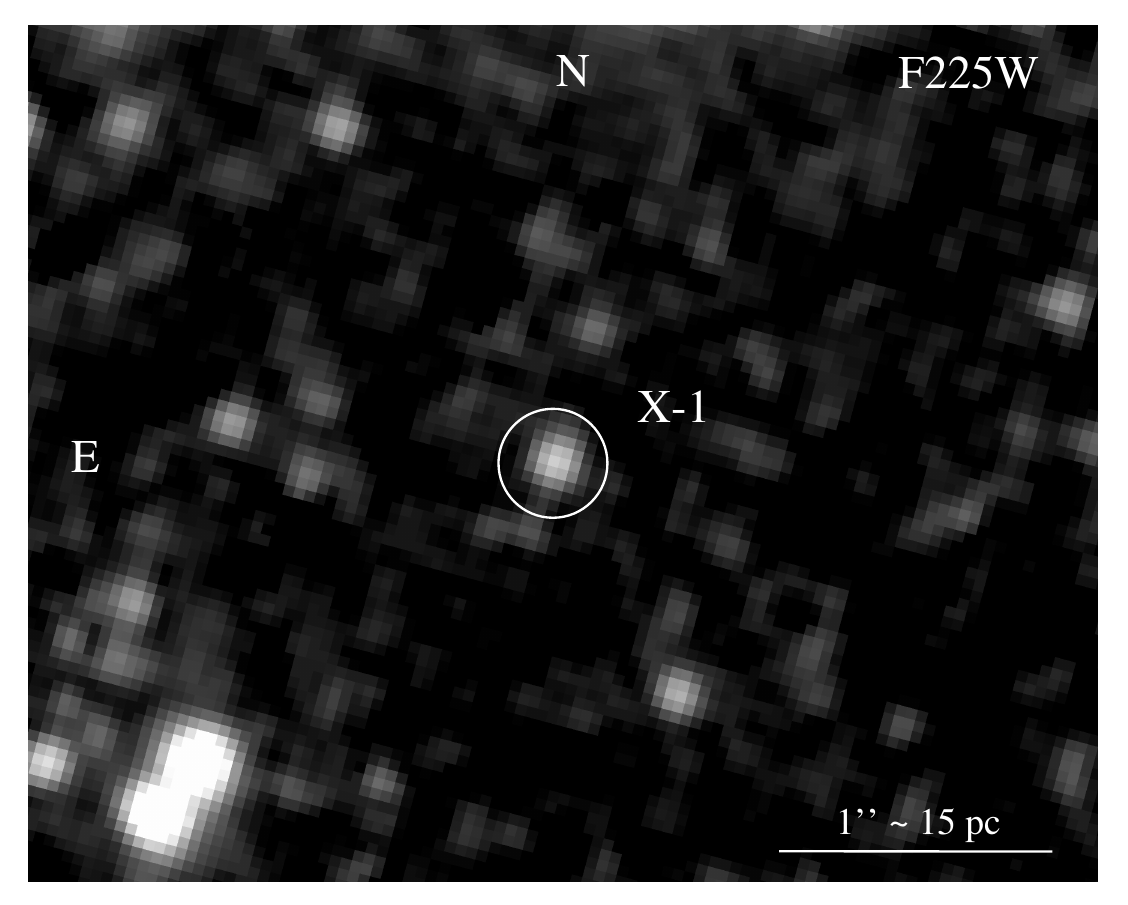}\\
    \vspace{-0.3cm}
    \includegraphics[width=1\linewidth]{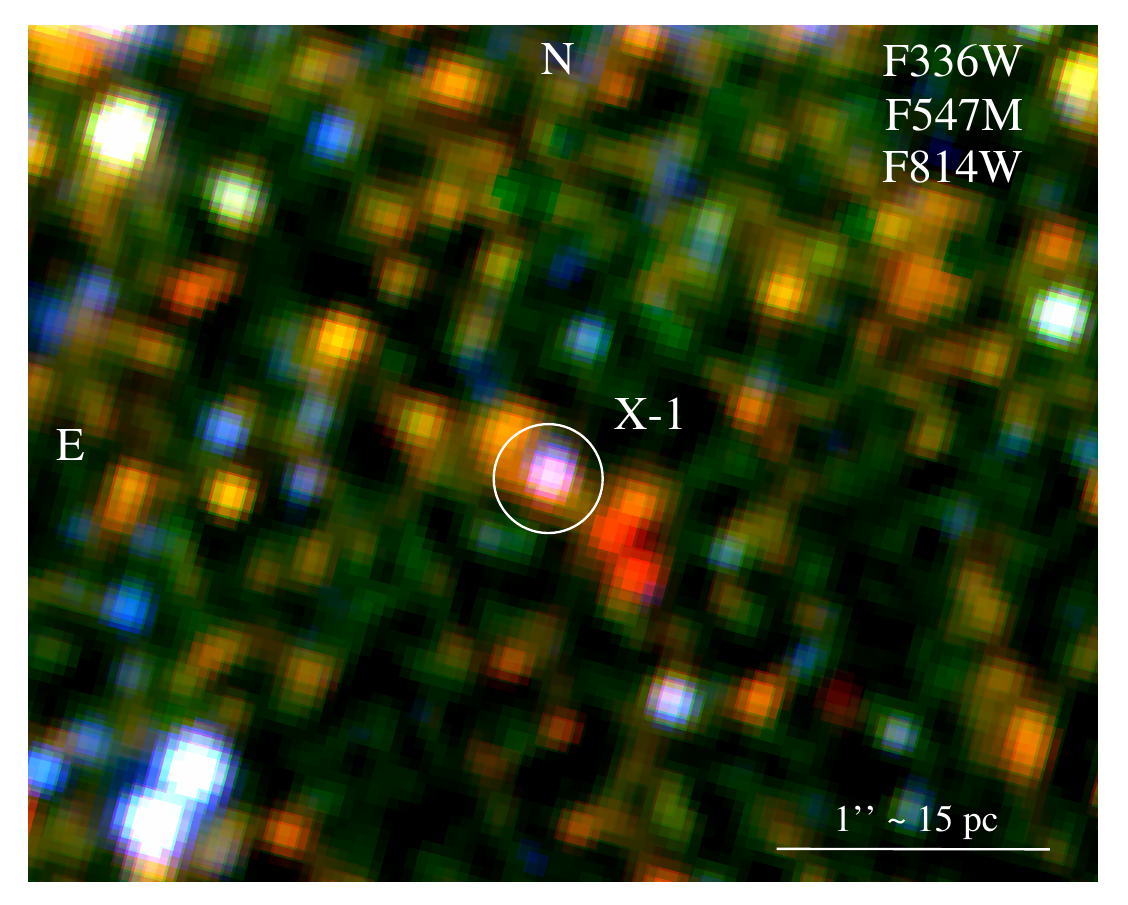}\\
    \vspace{-0.3cm}
    \includegraphics[width=1\linewidth]{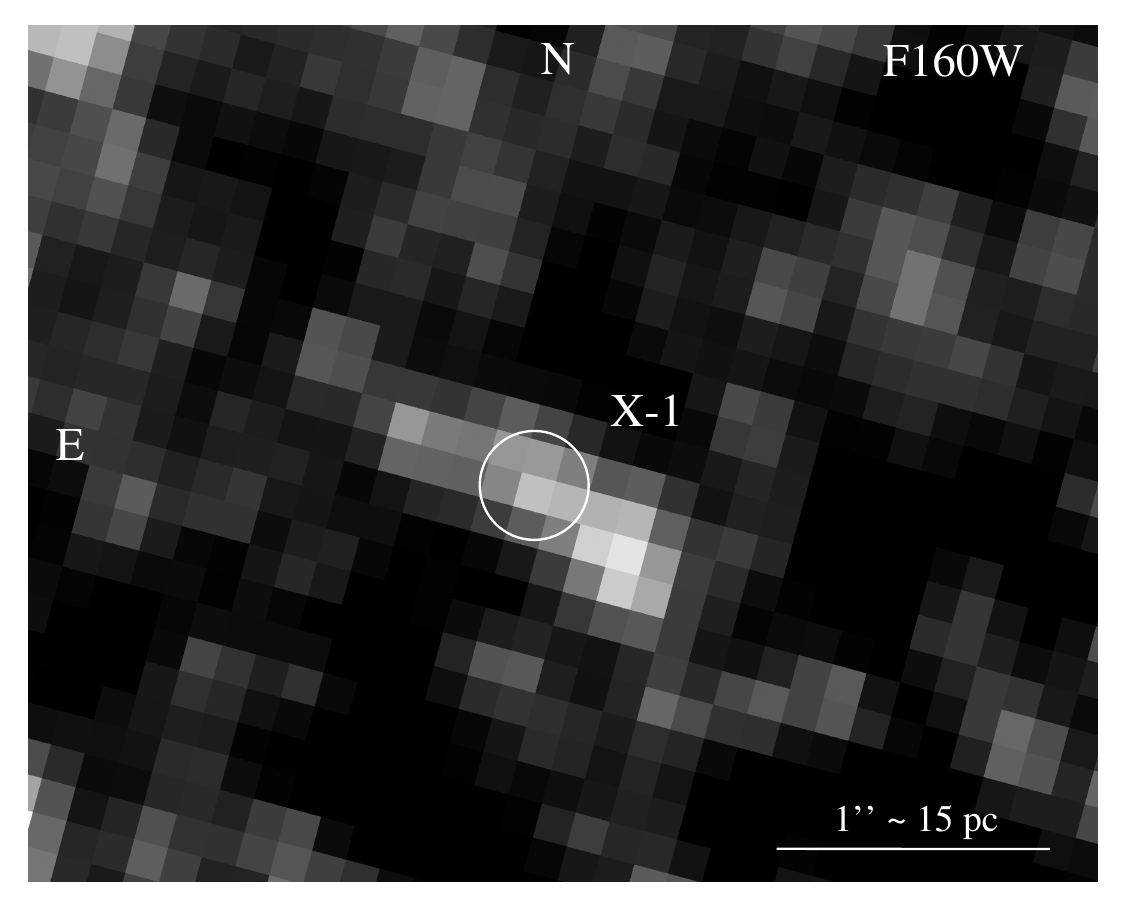}
    \caption{Top Panel: zoomed-in view of the field around X-1 in the near-UV band ({\it HST}/WFC3 UVIS filter F225W). The circle has a radius of 0\farcs{2}, approximately corresponding to the positional uncertainty of the X-ray source on the {\it HST} image. The image has been smoothed with a Gaussian kernel of radius 2 pixels. Middle panel: {\it HST}/WFC3 UVIS true-color image of the same field, also smoothed with a 2-pixel Gaussian. Red = F814W, green = F547M, blue = F336W. Bottom panel: (unsmoothed) {\it HST}/WFC3 IR image of the same field (F160W filter).}
    \label{fig2}
\end{figure}

\subsection{Previous studies of NGC\,4214 X-1}
NGC\,4214 (Figures \ref{fig1},\ref{fig1b}) is a dwarf starburst galaxy \citep[{\it e.g.},][]{mackenty00,hartwell04,ubeda07,mcquinn10,dopita10,williams11,choi20} at a distance of $\approx$($3.0 \pm 0.1)$ Mpc \citep{dalcanton09,dopita10}, corresponding to a distance modulus of $(27.4 \pm 0.1)$ mag. It has a current star formation rate $\approx$0.05--0.10 $M_{\odot}$ yr$^{-1}$ \citep{hartwell04,mcquinn10,williams11}, and 
an average metallicity similar to the Large Magellanic Cloud \citep{kobulnicky96,williams11}. Despite its obvious recent burst of star formation, only $\sim$1\% of the stellar mass formed in the last 100 Myr, and only $\sim$10\% in the last 4 Gyr \citep{williams11}; star formation activity appears to have picked up recently, especially over the last $\sim$10 Myr. Thus, its point-source X-ray population may include both HMXBs and low-mass X-ray binaries (LMXBs).

Previous X-ray studies of NGC\,4214 with the {\it Chandra X-ray Observatory} and {\it XMM-Newton} showed \citep{hartwell04,binder15} both diffuse emission from star-forming regions, and point sources (X-ray binaries and supernova remnants (SNRs)). Among the point-source population, the brightest source (CXOU J121538.2$+$361921 = X-1) reaches a peak luminosity of $\approx$10$^{39}$ erg s$^{-1}$ \citep{ghosh06}. The most remarkable property of X-1, discovered and discussed in details by \cite{ghosh06}, is an apparent X-ray eclipse, with a period of 3.62 hr (most likely, the binary period). From their X-ray timing and spectral analysis, \cite{ghosh06} argued that the most likely interpretation is a NS or BH accreting from a stripped He star via Roche lobe overflow. The observed binary period implies a donor-star radius $R_2 \sim 0.5$--$2  R_{\odot}$ for plausible ranges of primary masses and mass ratios. An intermediate-mass He star ($M_2 \lesssim 3$--$4 M_{\odot}$) was considered a more likely donor than a classical WR star, because \cite{ghosh06} found no evidence of strong winds (no significant changes in neutral absorption and spectral hardness ratio at eclipse ingress/egress). A low-mass main-sequence donor ($M_2 \sim 0.4 M_{\odot}$) would have been consistent with the short binary period but not with the apparently persistent nature of the X-ray source over several years. However, the non-detection of an optical counterpart made it impossible to provide stronger constraints.

\begin{deluxetable}{lccccr}[!]
\tablecaption{Log of the {\it Chandra}/ACIS observations of X-1. \textcolor{black}{Offset values indicate the off-axis distance of X-1 in each observation. Count rates are observation-averaged background-subtracted values in the 0.3--8 keV band.} \label{tab:OD}}
\tablewidth{0pt}
\tablehead{
\colhead{ObsID} & \colhead{Date} & \colhead{Start Time} & \colhead{Exp.} & \colhead{\textcolor{black}{Offset}} & \colhead{Count Rate}\\[-5pt]
 & & (MJD) & (ks) &  & ($10^{-3}$ ct s$^{-1}$)}
\startdata
2030  & 2001-10-16 &  52198.7957 & 26.4 & 1\farcm9 & $35.7 \pm 1.2$\\
4743  & 2005-04-07 &  53098.1970 & 27.2 & 2\farcm0 & $7.1 \pm 0.5$\\
5197  & 2005-08-03 &  53216.3293 & 28.6 & 1\farcm9 & $29.6 \pm 1.0$\\
22372 & 2020-08-01 &  59062.4143 & 59.3 & 0\farcm1 & $22.0 \pm 0.6$
\enddata
\end{deluxetable}

\begin{deluxetable}{ccccc}[!]
\tablecaption{{\it HST}/WFC3 observations of X-1 used for this study\label{tab:HST}. The dataset in the F275W filter was taken on 2022 November 14; all the others are from 2009 December 22--25.}
\tablewidth{0pt}
\tablehead{
\colhead{Filter} & \colhead{Exp} & \colhead{$\lambda_{\rm eff}$} & \colhead{Brightness} & \colhead{Flux Density}
\\[-5pt]
 & (s) & (\AA) & (Vegamag) & (erg cm$^{-2}$ s$^{-1}$ \AA$^{-1}$)
}
\startdata
F225W & 1665 & 2372.1 & $22.54\pm0.06$ & $(4.1\pm0.2) \times 10^{-18}$ \\ 
F275W & 2380 & 2709.7 & $22.75\pm0.08$ & $(3.0\pm0.2) \times 10^{-18}$ \\ 
F336W & 1683 & 3354.6 & $23.16\pm0.03$ & $(1.79\pm0.03) \times 10^{-18}$ \\
F438W & 1530 & 4325.7 & $24.50\pm0.05$ & $(1.07\pm0.05) \times 10^{-18}$ \\
F547M & 1050 & 5447.5 & $24.09\pm0.05$ & $(0.87\pm0.05) \times 10^{-18}$  \\
F657N & 1592 & 6566.6 & NA & $(0.76\pm0.11) \times 10^{-18}$ \\
F673N & 2940 & 6765.9 & NA & $(0.99\pm0.06) \times 10^{-18}$ \\
F814W & 1339 & 8039.1 & $22.44\pm0.05$ & $(1.20\pm0.05) \times 10^{-18}$ \\
F110W & 1197 & 11534.5 & $21.07\pm0.05$ & $(1.50\pm0.07) \times 10^{-18}$ \\
F160W & 2397 & 15369.2 & $20.1\pm0.1$ & $(1.3\pm0.1) \times 10^{-18}$ \\
\enddata
\end{deluxetable}

\begin{figure*}
    \centering
    \includegraphics[width=0.49\linewidth]{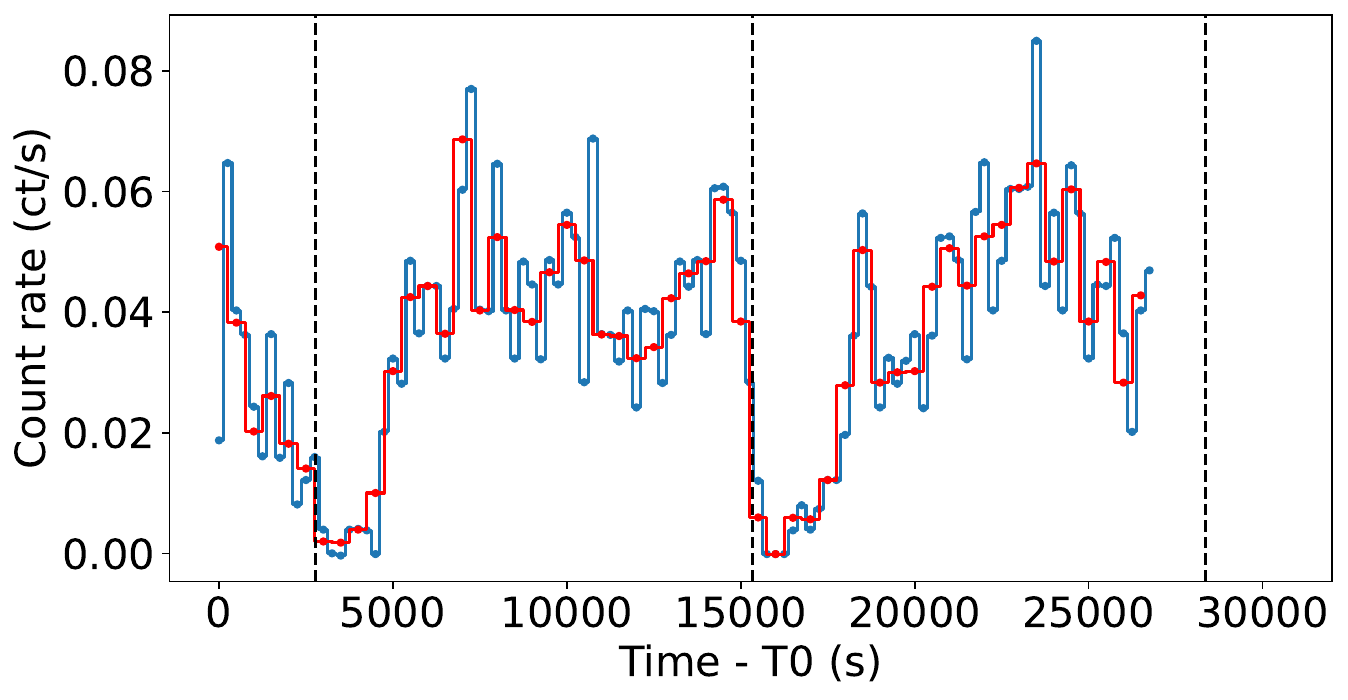}
    \includegraphics[width=0.49\linewidth]{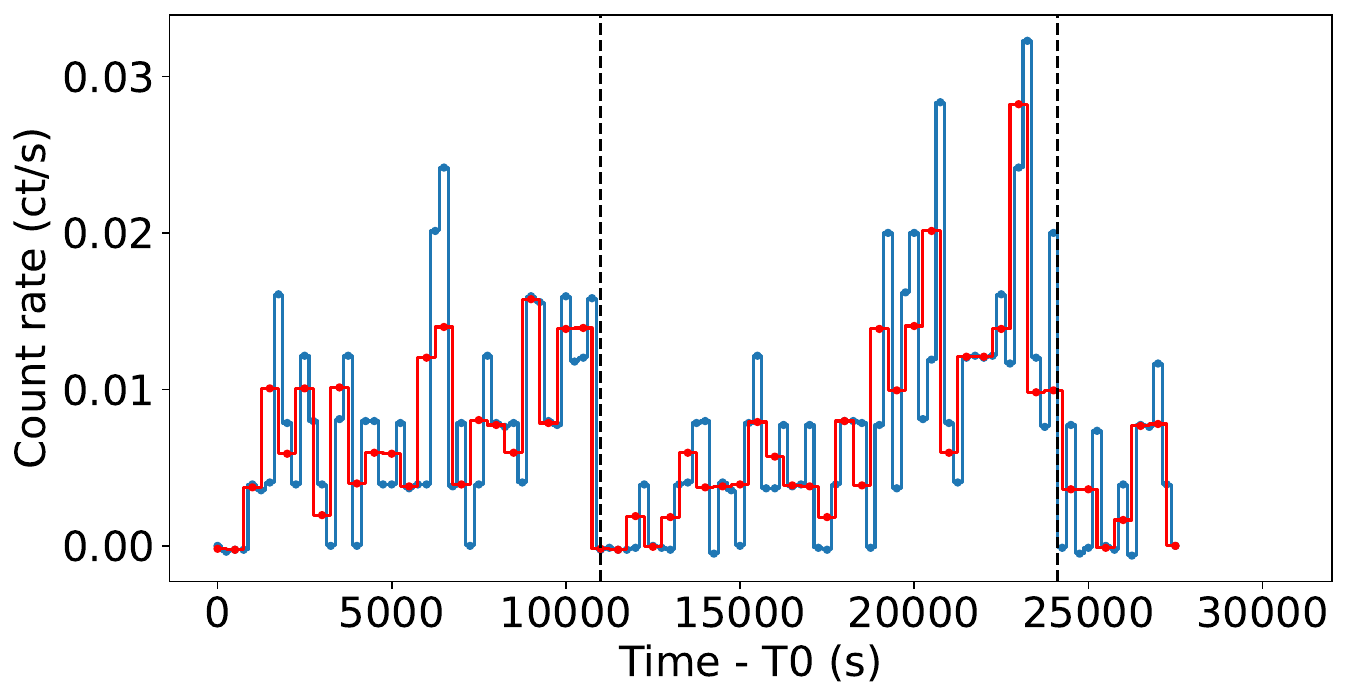}
    \includegraphics[width=0.49\linewidth]{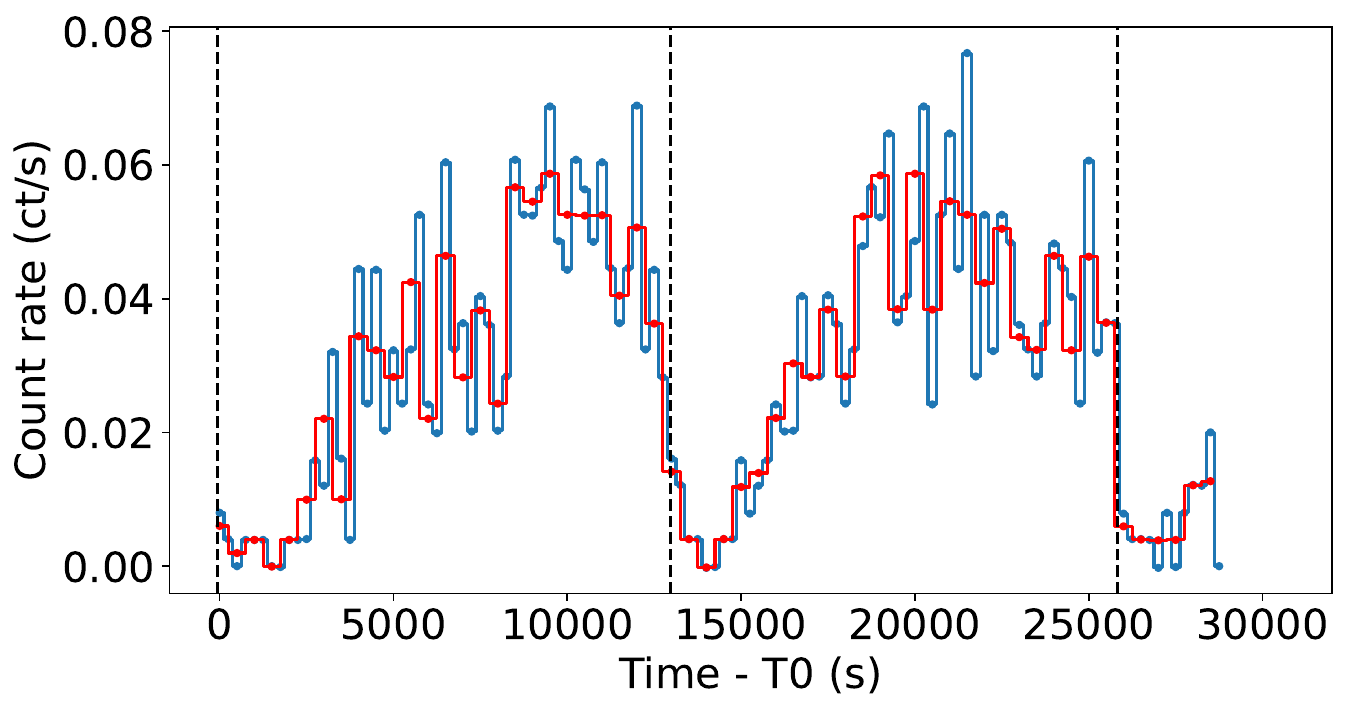}
    \includegraphics[width=0.49\linewidth]{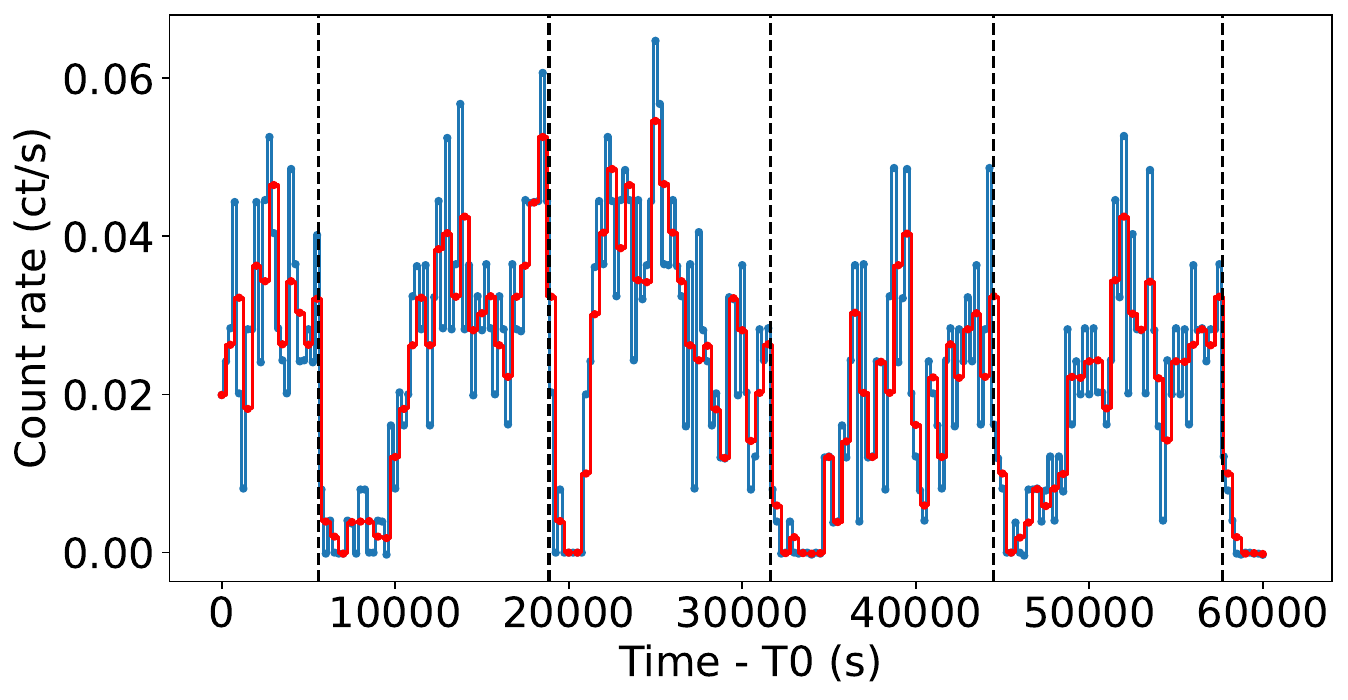}
    \caption{{\it Chandra}/ACIS light curves of X-1 in the 0.3--8 keV band. Top left: ObsIDs 2030; top right: ObsID 4743; bottom left: ObsIDs 5197; bottom right: ObsID 22372. Each light curve has been rebinned to 250 s (blue histogram) and 500 s (red histogram). Dashed black lines mark the 250 s bins corresponding to eclipse ingresses. The last dashed line in ObsID:2030 and first dashed line in ObsID:5197 are predicted eclipse times. In each plot, the reference time $T_0$ is the observation start time in Table \ref{tab:OD}. We have omitted the error bars for clarity. }
    \label{fig:eclipse_in.png}
\end{figure*}

\subsection{Specific objectives of our new study}
We carried out a follow-up optical/X-ray study of NGC\,4214 X-1. Our main specific objectives are: 
a) to identify a candidate optical counterpart in a set of {\it Hubble Space Telescope} ({\it HST}) observations that were not available at the time of the previous investigation, and verify whether the optical brightness is consistent with the suggested scenarios;
b) to determine whether the source is still in a X-ray bright state, two decades after the first {\it Chandra} observations, and/or whether there is evidence of state transitions; 
c) to determine the duration and average ingress/egress profiles of the X-ray eclipses more accurately; 
d) to determine whether the X-ray spectrum is dominated by a thermal (disk-blackbody) or a non-thermal (power-law) component.

To achieve those objectives, we have combined the three archival {\it Chandra} datasets from 2001--2005 with new data taken on 2020 August 1 (ObsID 22372, PI: R.~Soria). We have also used archival {\it HST} Wide Field Camera 3 (WFC3) images taken on 2009 December 22--25 (Program ID 11360, PI: R.~O’Connell) and 2022 November 14 (Program ID 17225, PI: D. Calzetti). 
A summary of the {\it Chandra} and {\it HST} observations is presented in Tables \ref{tab:OD} and \ref{tab:HST}. 

In Section \ref{sec:Chandra_HST}, we summarize our X-ray and optical data analysis procedures, and report on the improved astrometry, which enables the identification of a likely optical counterpart. In Sections \ref{sec:lc} and \ref{sec:spe}, we present the results of X-ray lightcurve and spectral modelling, respectively. In Section \ref{sec:hst}, we report on the optical brightness and colors. Finally, we discuss the implications of such results for the system parameters in Section \ref{sec:discussion}. 

\section{Methods} \label{sec:Chandra_HST}

\subsection{Chandra data analysis} 

In all four {\it Chandra} observations (Table \ref{tab:OD}), X-1 was located in the S3 chip of the Advanced CCD Imaging Spectrometer (ACIS) camera.
After downloading the data from the public archives, we reprocessed them with the Chandra Interactive Analysis of Observations ({\sc ciao}) software version 4.13 \citep{fruscione06}, Calibration Database 4.9.5. We created new level-2 event files with the {\sc ciao} tasks {\it chandra\_repro}. 
\textcolor{black}{
For each observation, we visually inspected the light curve of a large background region, binned at 250~s and 500~s time bins, to search for flares. We found an  $\sim$4~ks interval at the end of ObsID 4743 of strong flaring but determined that only a few extra background events (no more than 8.5\% of the total source$+$background counts) can be attributed to this activity, which occurred during an eclipse (hence, maximizing the background contribution to the total counts in the source region). We did not exclude this flare interval, preferring instead to retain the longest possible continuous time intervals to better conduct our searches for periodic signals.
}

We applied a barycenter correction with {\it axbary}, for timing analysis. We used {\it reproject\_obs} to create stacked images, {\it dmextract} to obtain lightcurves from each observation, and {\it srcflux} for preliminary count rate and flux estimates. We then used {\it specextract} to create spectra and associated response and ancillary response files. 
For timing and spectral analysis, we used source extraction circles of $3^{\prime\prime}$ radius, \textcolor{black}{centered using the {\it centroid} command in {\sc ds9}}, in each of the four observations. 
\textcolor{black}{Local background annuli were centered on the same coordinates as the source region with an inner radius of $4^{\prime\prime}$ and an outer radius of at least $17^{\prime\prime}$.}

We used NASA's High Energy Astrophysics Science Archive Research Center (HEASARC) software for further data analysis; in particular, the {\sc ftools}\footnote{http://heasarc.gsfc.nasa.gov/ftools} package \citep{blackburn95,heasoft14}. With {\sc ftools}'s {\it grppha} task, we regrouped the spectra to a minimum number of counts per bin, suitable for spectral fitting: $\geq$1 count per bin for the Cash statistics \citep{cash79} and $\geq$15 counts per bin for the $\chi^2$ statistics. 
\textcolor{black}{We fitted the spectra with {\sc xspec}, version 12.12.0 \citep{arnaud96}, over the standard 0.3--8 keV ACIS energy range. We tested all the models used in this study both on the minimally grouped spectra with the {\tt cstat} fit statistic \citep{cash79}, and on the more grouped version of the same spectra with the $\chi^2$ statistics, whenever possible ({\it i.e.}, for spectra with enough counts). In all those cases, our results are consistent between the two fitting statistics, as expected because of the asymptotic equivalence of the Poisson and Gaussian statistics. The spectrum from ObsID 4743 (Section 4.1) and some of the low-flux-threshold spectra (Section 4.2) do not have enough counts for $\chi^2$ fitting ({\it i.e.}, the regrouped spectra have only $\sim$10 bins and substantially undersample the spectral resolution, with a loss of information). Such spectra are better fitted with the {\tt cstat} statistic. For consistency, we report all the best-fitting parameter values, fluxes and confidence intervals obtained with {\tt cstat}.  Model absorbed and unabsorbed fluxes were obtained with the {\tt cflux} convolution model. Uncertainties for one interesting parameter (calculated with the {\it steppar} command) are reported at the confidence interval of $\Delta C = \pm$2.70: this is asymptotically equivalent to the 90\% confidence interval in the $\chi^2$ statistics.} 

\subsection{HST data analysis and refined astrometry}  \label{sec:HST_data}
NGC\,4214 was observed by {\it HST}/WFC3 on 2009 December 22--25, in several broad-band and narrow-band filters; those used for this study are listed in Table \ref{tab:HST}. An additional {\it HST}/WFC3 observation in the F275W filter was taken on 2022 November 14.
We downloaded the drizzled, calibrated {\it HST}/WFC3 image files 
(.drc files for WFC3-UVIS, .drz files for WFC3-IR) from the \textcolor{black}{Mikulski Archive for Space Telescopes\footnote{https://mast.stsci.edu/search/ui/{\#}/hst}}. We used the {\sc ds9} analysis tools for astrometry and photometry. Comparing the {\it Chandra} and {\it HST} images, we found at least nine X-ray sources with likely matches on the {\it HST} images. Six of them (Figure \ref{fig1}) are supernova remnants previously identified by \cite{dopita10} in the narrow-band images\footnote{An investigation of the X-ray properties of those SNRs is beyond the scope of this paper. However, we did measure their observed fluxes and intrinsic luminosities, and report the results in Appendix A.}.  They are extended in the {\it HST} images, and barely above the detection limit in the {\it Chandra} images. Thus, those sources are not useful to improve the relative astrometry of {\it Chandra} and {\it HST} images beyond an $\approx$0$\farcs{5}$ uncertainty. Another three {\it Chandra} sources correspond to point-like sources in the {\it HST} broad-band images. 
For those three associations, we measured an average offset of $0\farcs{14}$ and a maximum offset of $0\farcs{21}$. 
\textcolor{black}{
We applied a simple translation of the {\it Chandra} wcs coordinate system using the {\sc ciao} utility {\tt wcs\_update} to align the {\it Chandra} data to the {\it HST} coordinates. We do not have enough {\it Chandra}/{\it HST} associations to justify more complex transformations such as rotation and scaling. As for the {\it HST} absolute astrometry, we did not detect any systematic offset between matching sources in {\it HST}, {\it SDSS}, and {\it Gaia}. }

\textcolor{black}{Combining the uncertainty in the centroid of the X-ray source on the ACIS chip with the uncertainty in the relative astrometric alignment of {\it Chandra} and {\it HST}, we estimate a 90\% confidence radius of $\approx$0$\farcs{2}$ for the location of X-1 on the {\it HST} images.} We find that there is only one point-like optical candidate in the {\it Chandra} error circle of X-1 (in fact, right at the centre).  The candidate point-like optical counterpart is also the brightest blue object within $\approx$30 pc of the X-ray position (Figure \ref{fig2}). In summary, the coordinates of X-1 are R.A.(J2000) $= 12^{h}15^{m}38^{s}.25 (\pm 0^{s}.01)$, Dec(J2000) $= +36^{\circ} 19^{\prime} 21\farcs{0} (\pm 0\farcs{1})$.  

To estimate the flux from the candidate {\it HST} counterpart, \textcolor{black}{we used standard {\sc ds9} aperture photometry tools.} Specifically, we defined a source circle of radius $0\farcs{16}$ for the WFC3-UVIS images (optical/UV) and $0\farcs{20}$ for the WFC3-IR images (near-infrared). The circle was centred on the optical centroid of the {\it HST} counterpart \textcolor{black}{(estimated with the {\it centroid} tool in {\sc ds9}}); that position is also within a single WFC3-UVIS pixel of the {\it Chandra} position. 
\textcolor{black}{
In each image, we selected and averaged three local background regions with a combined area of about twelve times the source circle. Such regions were suitably placed to avoid the bright stars north-east and south-west of X-1. We repeated the background estimation several times for each image, with different placements and relative sizes of the background regions, and took the median value of the aperture-limited net count rate. The dispersion in the net count rate values for different background locations gave us an estimate of the uncertainty, more comprehensive than an estimate based on Poisson statistics of the number of detected electrons. We converted aperture-limited count rates to infinite-aperture values $\rm{CR_{inf}}$, using the online tables of encircled energy fractions\footnote{WFC3 Instrument Handbook, Chapter 6 for UVIS and Chapter 7 for IR; https://hst-docs.stsci.edu/wfc3ihb.}. 
We then obtained magnitudes and flux densities from the infinite-aperture count rates, using the latest tables of UVIS zeropoints\footnote{http://www.stsci.edu/hst/wfc3/analysis/uvis\_zpts/uvis1\_infinite} and IR zeropoints\footnote{http://www.stsci.edu/hst/wfc3/ir\_phot\_zpt}, with the standard relation
\begin{equation}
    {m_{\rm vega}} = -2.5\times\log_{10}(\rm{CR_{inf}}) + \rm{ZP_{vega}},
\end{equation}
where $m_{\rm vega}$ is the apparent magnitude in the VegaMag system and $\rm{ZP_{vega}}$ is the corresponding zeropoint. Finally, we converted each WFC3 datapoint (in flux density units) to pha $+$ rsp file format, suitable to {\sc xspec} modelling. This is a standard technique that allows joint fitting of X-ray and optical data. For this, we used the {\sc ftools} task {\it ftflx2xsp}\footnote{https://heasarc.gsfc.nasa.gov/lheasoft/ftools/headas/ftflx2xsp.html}.}

\section{X-ray timing results}\label{sec:lc}

\begin{figure}[!]
  \centering
    \includegraphics[width=\linewidth]{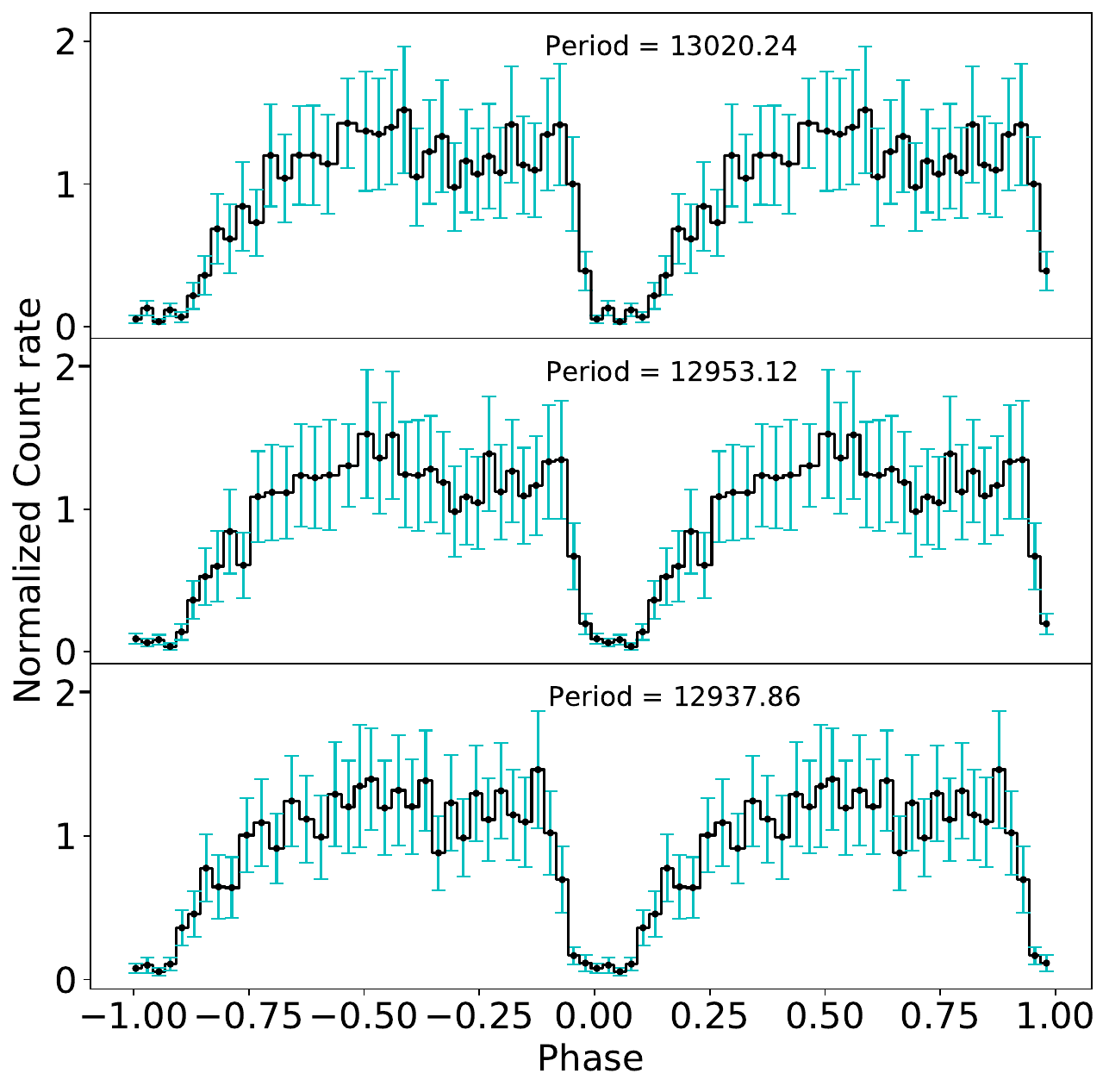}
    \caption{Average {\it Chandra} light-curve of X-1 in the 0.3--8 keV band, folded on three alternative and statistically equivalent binary periods, and rebinned to 35 phase intervals per period. Top panel: folding period of 13020.24 s; middle panel: folding period of 12953.12 s; bottom panel: folding period of 12937.86 s.}
    \label{fig:phase_count_6p}
\end{figure}

\subsection{Period searching}

Before searching for the best-fitting period, we renormalized the {\it Chandra} light curves of X-1 to the average net count rate in each observation. We did this to correct for the decreasing sensitivity of ACIS-S in more recent epochs, which would have biased the period search in favour of the earlier observations.

First, we applied {\sc ftools}'s {\it efsearch} task to the full light curve (all four epochs). The best-fitting period depends both on the time binning of the light curve, and on the time resolution used for period searches. The search is also hampered by the large gaps between epochs. We folded several candidate best-fitting solutions with the {\it efold} task and visually inspected them. We find that the most stable solution (obtained with 500 s binning of the light curves and 0.01 s period search resolution) is at $P \approx 13020.24$ s. The corresponding folded light curves is shown in the top panel of Figure \ref{fig:phase_count_6p}.

To test this value and perhaps obtain a more robust estimate, we took advantage of the 11 X-ray eclipses covered by our observations (Figure \ref{fig:eclipse_in.png}).
At all four epochs, eclipse ingresses are typically sharper than the corresponding egresses. Thus, we used ingress times to determine the best-fitting period. For each eclipse, we defined the ingress time $t_e$ as the mid-point of the 250 s bin with the largest count-rate drop, and attribute an uncertainty of $\pm125$ s to that datapoint, for period searching.

We used a least-squares algorithm to fit a linear function 
to the cycle numbers and times of eclipse ingress:
\begin{align}\label{equ:cycle_period}
    t_e = I\times P + T_0,    
\end{align}

where $I$ is the cycle number, $P$ is the orbital period, and $T_0$ is a reference eclipse ingress time. We obtained a best-fitting period $P = 12937.86\pm267.21$ s. Of the four {\it Chandra} epochs, ObsID 4743 is when the source was faintest, and it is more difficult to pinpoint the ingress times of its two eclipses with the same confidence as for the other 9 events. To test whether this may affect our period search, we repeated our fitting without ObsID 4743. We obtained a period $P=12953.12\pm245.98$ s.
The $\sim$$\pm$250 s error range of the best-fitting period is the envelope within which we find acceptable folding solutions; it does not mean that any period within that time range is equally likely. Instead, there is a discrete list of acceptable periods, each of them corresponding to a different, integer number of orbital cycles elapsed from the first to the last observation. Given that the observations span 20 years, and there are gaps of many years between them, it is unfortunately impossible to unequivocally phase-connect the light-curves.

\subsection{Eclipse Profile}

Regardless of the choice of folding period, the folded eclipse profile has a faster (though not instantaneous) ingress and a slower egress, lasting as long as $\sim$0.3 times the binary period before the flux returns to the out-of-eclipse level. The eclipse itself is well defined in the average light-curve (Figure \ref{fig:phase_count_6p}) although it can be difficult to define in individual observations, because of the difficulty to determine when the true eclipse ends and the slow egress begins. This ``fuzziness'' is usually a sign that the eclipsing donor star is surrounded by a dense wind; we will discuss this issue later. 

Based on the average light-curve, we estimate an eclipse duration $\Delta T_{\rm ecl} \approx 2050 \pm 175$ s. Among the individual eclipses observed in the four epochs, the one with the sharpest ingress and egress (which we interpret as the epoch in which the effect of the wind was least important) was the second eclipse in ObsID 22372: in that case, too, the duration was $\approx$2000 s. Thus, we shall use this average value for an estimate of the size of the donor star. Correspondingly, the eclipse fraction $\Delta_{\rm ecl} \equiv \Delta T_{\rm ecl}/P = 0.158 \pm 0.014$.

\begin{deluxetable*}{lcccc}
\tablecaption{Best-fitting absorbed power-law model of the {\it Chandra}/ACIS spectra of X-1 in the four epochs. The spectra were fitted with the Cash statistics. The Galactic column density was fixed at $N_{\rm {H,Gal}} = 4 \times 10^{20}$ cm$^{-2}$. Error ranges are 90\% confidence limits for one interesting parameter.}
\tablewidth{0pt}
\tablehead{
\colhead{Model Parameters} & \multicolumn{4}{c}{Values in Each ObsID}
\\
 &   2030 & 4743 & 5197 & 22372
}
\startdata
  & & & & \\[-8pt]
   $N_{\rm {H,int}}$   ($10^{21}$ cm$^{-2}$)   &  $ 1.4^{+0.5}_{-0.5}$ & $<1.2$ & $2.0^{+0.7}_{-0.6}$ & $2.5^{+1.3}_{-1.2}$  \\[4pt]
   $\Gamma$      &  $1.77^{+0.16}_{-0.15}$ & $2.19^{+0.41}_{-0.27}$ & $1.84^{+0.17}_{-0.16}$ & $1.95^{+0.16}_{-0.15}$ \\[4pt]
   $N_{\rm {po}}^a $ 
              &  $5.8^{+0.9}_{-0.7}$ & $1.1^{+0.4}_{-0.2}$ & $5.8^{+0.9}_{-0.9}$ & $8.5^{+1.6}_{-1.4}$ \\[4pt]
   C-stat/dof     &      $229.1/331$ (0.69) & $86.1/115$ (0.75) & $193.0/261$ (0.74) & $327.2/322$ (1.02)\\[4pt]
   $f_{0.3-8}$ ($10^{-13}$ erg cm$^{-2}$ s$^{-1}$)$^b$ 
       & $2.7^{+0.3}_{-0.2}$   
       & $0.46^{+0.08}_{-0.08}$
       & $2.4^{+0.3}_{-0.2}$ 
       & $3.1^{+0.2}_{-0.2}$  \\[4pt]
\enddata
\tablecomments{
$^a$: units of $10^{-5}$ photons keV$^{-1}$ cm$^{-2}$ s$^{-1}$ at 1 keV.\\
$^b$: observed flux in the 0.3--8 keV band\\
}
\label{tab:powerlaw_result}
\end{deluxetable*}

\begin{figure}
\hspace{-0.5cm}
    \includegraphics[height=1.05\linewidth, angle=270]{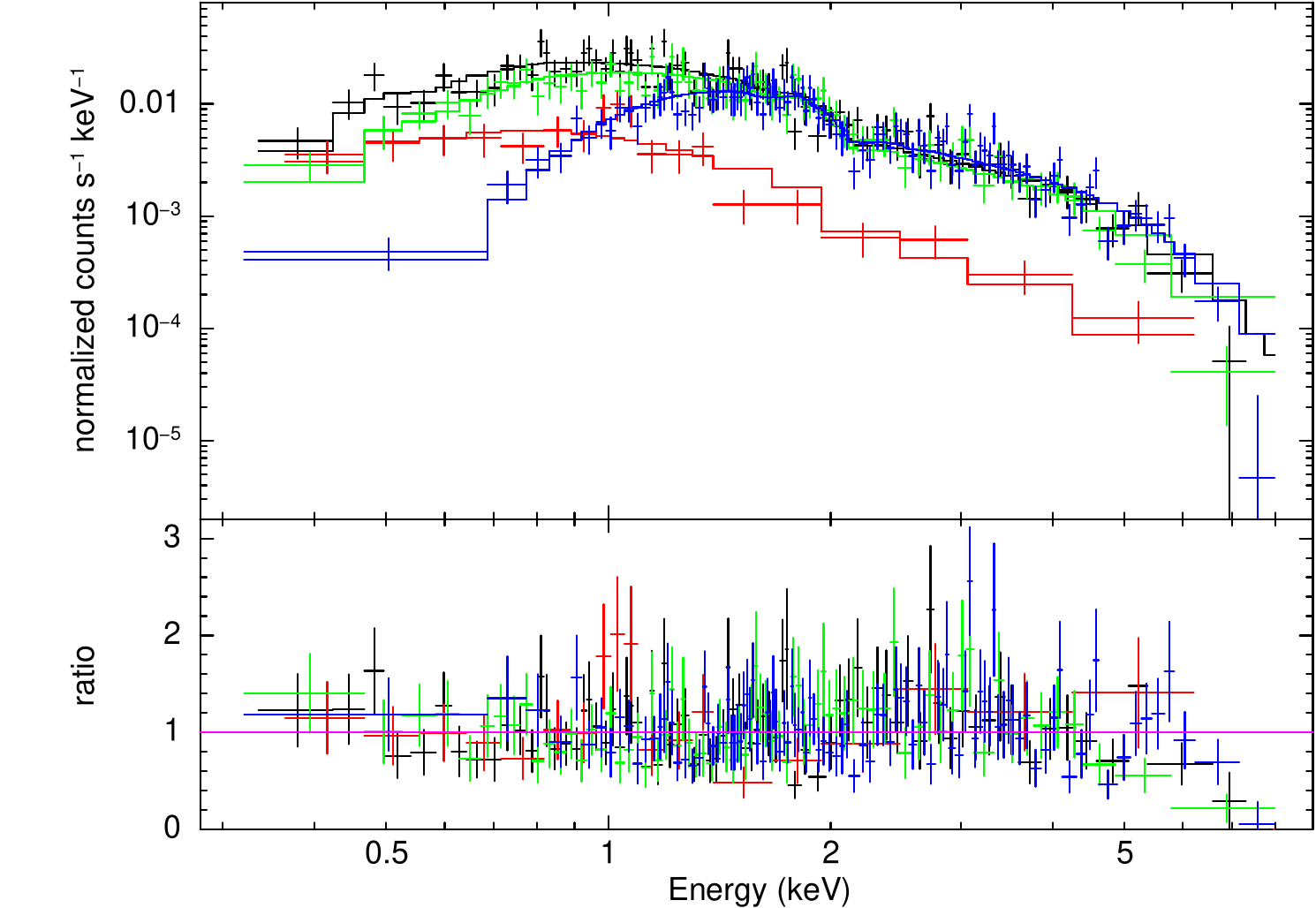}
    \caption{{\it Chandra}/ACIS-S spectra of X-1 in the four epochs of observations, with best-fitting power-law model and data/model ratios. Black datapoints and model histogram are for ObsID 2030; red for ObsID 4743; green for ObsID 5197; blue for ObsID 22372. See Table \ref{tab:powerlaw_result} for the fit parameters. The spectra were fitted with the Cash statistics; the datapoints have been rebinned to a signal-to-noise ratio $>$3 for plotting purposes only. The decreasing count rates below 1 keV from the black to the green and blue spectra are the result of the decrease of ACIS-S sensitivity from 2001 to 2020, for the same absorbing column density.}
    \label{fig:four_spectra}
\end{figure}

\begin{figure}
\hspace{-0.5cm}
    \includegraphics[height=1.05\linewidth, angle=270]{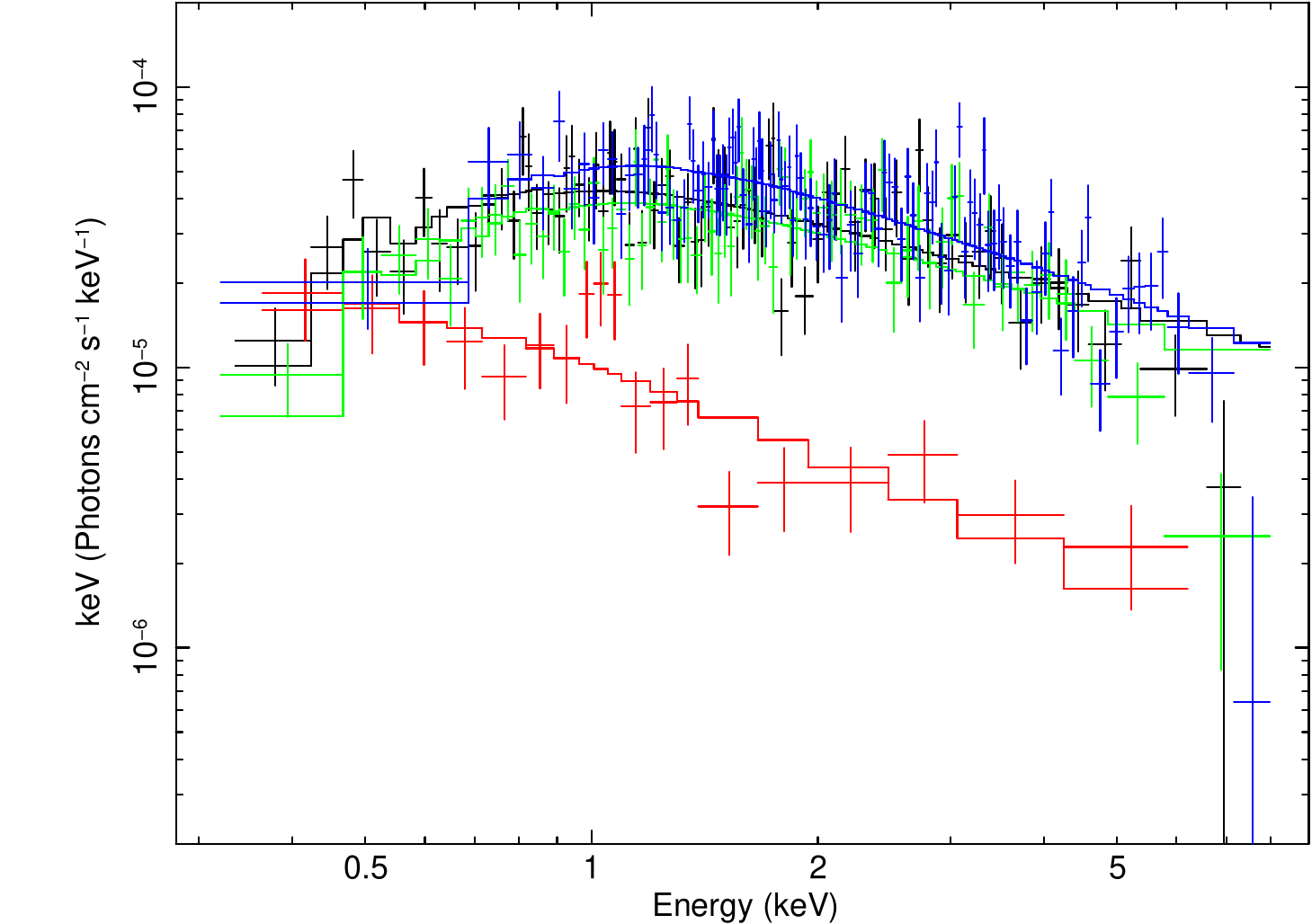}
    \caption{Unfolded {\it Chandra}/ACIS-S spectra of X-1 in the four epochs of observations, based on a power-law model (Table \ref{tab:powerlaw_result}). Color coding and re-binning are as in Figure \ref{fig:four_spectra}. The unfolded spectra highlight that X-1 was in a different state in ObsID 4743, compared with the other three epochs.}
    \label{fig:four_spectra_unfolded}
\end{figure}

\section{X-ray spectral results}\label{sec:spe}

\subsection{Thermal and non-thermal components}

To begin, we compared the four spectra corresponding to the four {\it Chandra} observations, without any other selection criteria. We used a simple power-law model, absorbed with a Galactic line-of-sight component plus an intrinsic component ({\it tbabs}$_{\rm Gal}$ $\times$ {\it tbabs}$_{\rm int}$ $\times$ {\it pow}). The foreground column density was fixed at $N_{\rm H} = 4 \times 10^{20}\;\rm{cm^{-2}}$ \citep{HI4PI16}, while the intrinsic column was left as a free parameter.
The purpose of this initial exercise was simply to assess whether the source was in a similar spectral state during the four epochs. We found that an absorbed power-law is a decent fit for each of the four epochs (Figure \ref{fig:four_spectra} and Table \ref{tab:powerlaw_result}); there are hints of a systematic curvature and high-energy turnover in the residuals, but they are not statistically significant in individual spectra, given their moderately low signal to noise level. In three of the four epochs, the observed spectra have similar slopes and normalizations (Figure \ref{fig:four_spectra}); the difference in the observed spectral shapes at low energies is only due to the loss of sensitivity of the ACIS-S detector over the years. The only epoch with a clearly different spectral shape is ObsID 4743 (red data points and model in Figures \ref{fig:four_spectra}, \ref{fig:four_spectra_unfolded}), with a softer power-law slope and lower luminosity. 

\begin{figure}
\hspace{-0.5cm}
    \includegraphics[height=1.05\linewidth, angle=270]{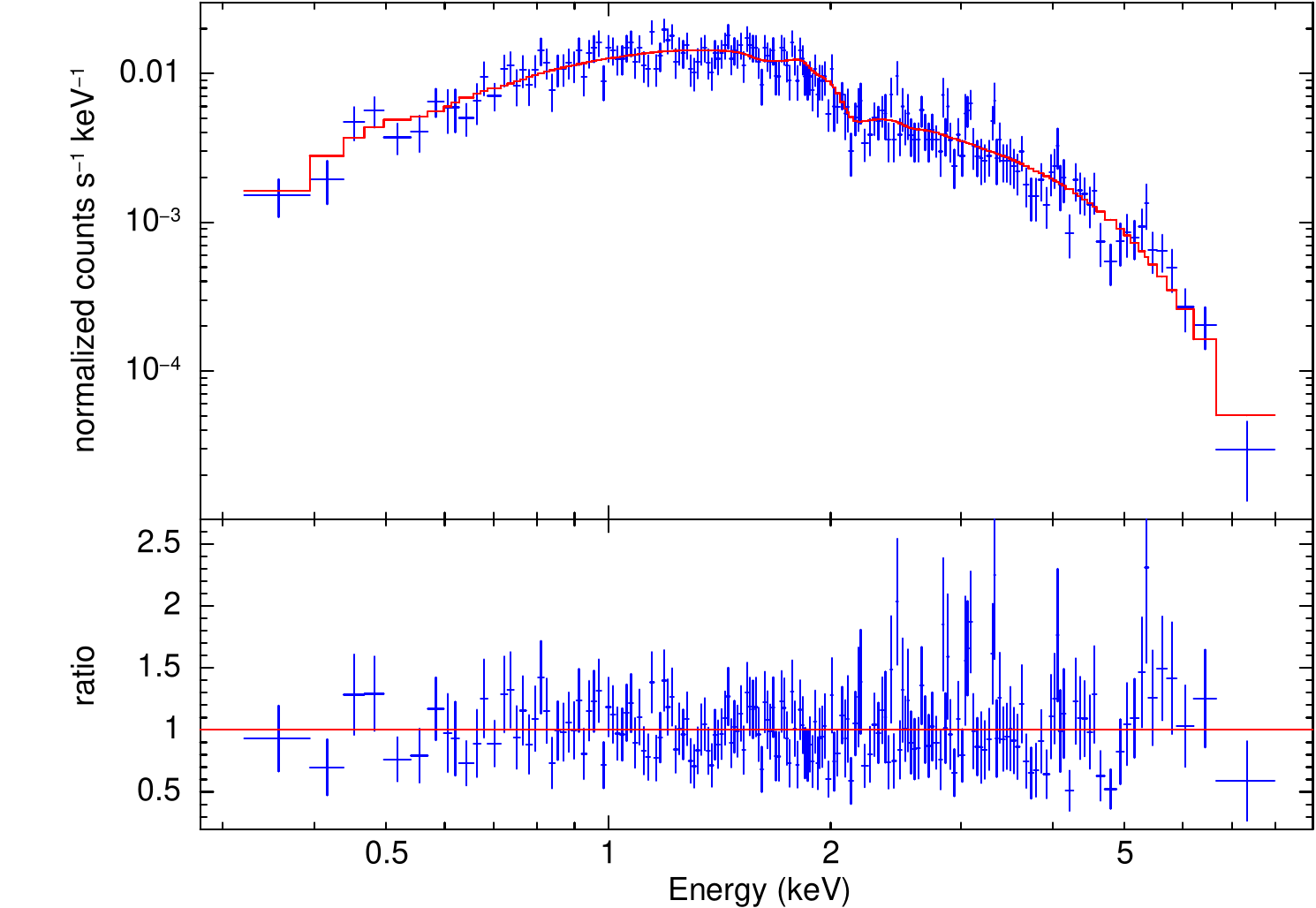}
    \caption{Combined {\it Chandra}/ACIS-S spectrum of X-1 from ObsIDs 2030, 5197 and 22372, with best-fitting {\it diskbb} model and data/model ratios.  See Table \ref{tab:comb_result} for the fit parameters. The spectra were fitted with the Cash statistics; the datapoints have been rebinned to a signal-to-noise ratio $>$3 for plotting purposes only.}
    \label{fig:combined_diskbb}
\end{figure}

Based on those preliminary results, we built a combined spectrum of ObsIDs 2030, 5197 and 22372 (leaving out ObsID 4743, as explained above). This was done with the {\tt combine\_spectra} script in the {\sc ciao} tool {\sl specextract}, which creates exposure-weighted {\it arf} files, and {\it arf}-weighted response files.
We found that an absorbed disk-blackbody model (Figure \ref{fig:combined_diskbb}, Table \ref{tab:comb_result}) is a better fit than an absorbed power-law (C-statistics of 338.8 for 380 degrees of freedom for the disk model, compared with 375.4/380 for the power-law model).
\footnote{\textcolor{black}{One possible bias in the combine procedure arises when some spectra have significantly higher background level than others; however, in our case, all three input spectra have very low background level. Alternatively, we fitted the three individual spectra simultaneously with all parameters linked. The resulting best-fitting parameter values are fully consistent with those found from a fitting of the combined spectrum.}}
The inner-disk radius $R_{\rm in}$, estimated from the disk normalization \citep{kubota98}, depends on the viewing angle $\theta$. Given the presence of eclipses, we know that the binary orbital plane is seen at high inclination, $i \ga 70^{\circ}$; we assume that the X-ray emitting part of the disk is aligned with the orbital plane ($\theta \approx i$), in the absence of specific evidence against this simplified scenario. For example, for $\theta = 75^{\circ}$, $R_{\rm in} \approx 45$ km, and for $\theta = 80^{\circ}$, $R_{\rm in} \approx 55$ km (Table \ref{tab:comb_result}). These are typical values found in stellar-mass BH X-ray binaries in the high/soft state \citep{mcclintock14,remillard06}.  For disk emission, the relation between flux and luminosity is also a function of viewing angle: $L_{\rm X} \approx 2\pi d^2 f_{\rm X}/(\cos \theta)$. For $\theta = 75^{\circ}$, we estimate a 0.3--10 keV luminosity $L_{\rm X} \approx 5.8 \times 10^{38}$ erg s$^{-1}$, and for $\theta = 80^{\circ}$, $L_{\rm X} \approx 8.6 \times 10^{38}$ erg s$^{-1}$ km. Considering that such values include part of the exposure time in which the source was in eclipse ingress/egress, we conclude that the intrinsic X-ray luminosity out of eclipse is close to $\approx$10$^{39}$ erg s$^{-1}$, Eddington limit of a stellar-mass BH. The best-fitting peak color temperature $kT_{\rm in} \approx (1.3 \pm 0.1)$ keV is also consistent with the temperatures observed at the peak of the high/soft state (near their Eddington limit) in Galactic stellar-mass BH transients \citep{kubota04,abe05,sutton17}.

Next, we tried more complex fitting models for the combined three-ObsID spectrum: i) a standard disk-blackbody plus power-law model; \textcolor{black}{ii) a 
disk model with a free power law dependence of the color temperature on radius, $T \propto R^{-p}$ ({\it diskpbb} in {\sc xspec}, generally used as a simple approximation of super-critical slim-disk models)}; iii) a Comptonization model ({\it simpl} $\times$ {\it diskbb}). None of those models provides a statistically significant improvement to the {\it diskbb} model. Finally, we tried adding an ionized absorber ({\it absori} model) but found no statistical need for it. When the ionization parameter in {\it absori} is frozen to its maximum value of  1000, we can only determine an upper limit to the possible ionized column density, $N_{\rm H}^{\rm ion} < 3.3 \times 10^{22}$ cm$^{-2}$ at the 90\% confidence level.

\begin{figure}
\hspace{-0.5cm}
    \includegraphics[height=1.05\linewidth, angle=270]{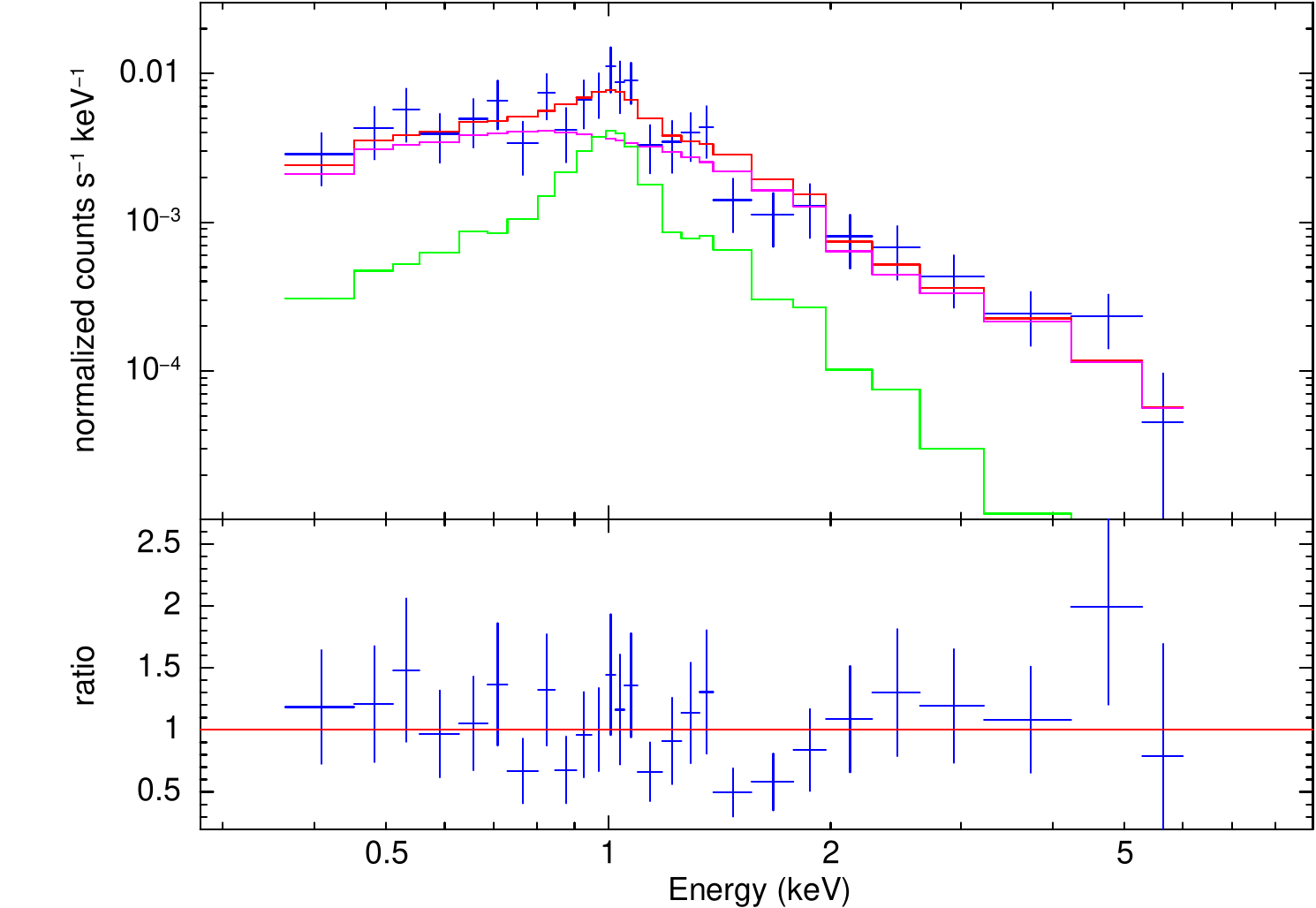}
    \caption{{\it Chandra}/ACIS-S spectrum of X-1 from ObsID 4743, with best-fitting {\it pow} $+$ {\it apec} model and data/model ratios.  See Table \ref{tab:4743_result} for the fit parameters. The spectra were fitted with the Cash statistics; the datapoints have been rebinned to a signal-to-noise ratio $>$3 for plotting purposes only. The green histogram shows the contribution of the {\it apec} component; the magenta histogram represents the power-law component; the red histogram shows the combined model.}
    \label{fig:4743_apec}
\end{figure}

We modelled the spectrum from ObsID 4743 separately. We tried adding an optically thin thermal plasma \textcolor{black}{({\it apec})} component (Figure \ref{fig:4743_apec} and Table \ref{tab:4743_result}) to the pure power law model, justified by the residuals seen around 1 keV. Using the task {\it simftest} in {\sc xspec}, we found a statistical improvement at the 99\% confidence level over the simple power-law model (C-statistics of 74.9 for 113 degrees of freedom, and  83.0/115, respectively). Conversely, we verified that the same {\it apec} component would not be statistically significant in the other three epochs, because of their higher continuum flux.

\begin{deluxetable}{lc}
\tablecaption{Best-fitting parameters of the combined {\it Chandra}/ACIS spectrum from ObsIDs 2030, 5197, 22372, fitted with the Cash statistics. The model is {\it tbabs}$_{\rm Gal}$ $\times$ {\it tbabs}$_{\rm int}$ $\times$ {\it diskbb}. The Galactic column density was fixed at $N_{\rm {H,Gal}} = 4 \times 10^{20}$ cm$^{-2}$. Error ranges are 90\% confidence limits for one interesting parameter.}
\tablewidth{0pt}
\tablehead{
\colhead{Model Parameters} & \colhead{Values}
}
\startdata
   &   \\[-8pt]
$N_{\rm {H,int}}$   ($10^{20}$ cm$^{-2}$)   &  $ <2.9$  \\[4pt]
   $kT_{\rm in}$ (keV)     &  $ 1.33^{+0.07}_{-0.07}$\\[4pt] 
   $N_{\rm {dbb}} $  ($10^{-3}$ km$^2$)$^a$ &  $ 4.3^{+0.9}_{-0.7}$\\[4pt]
   $R_{\rm {in}}\sqrt{\cos \theta} $  (km)$^b$ & $23.5^{+2.4}_{-2.1}$\\[4pt]
   C-stat/dof     &   $338.8/380$ (0.89)\\[4pt]
   $f_{0.3-8}$ ($10^{-13}$ erg cm$^{-2}$ s$^{-1}$)$^c$ 
       & $2.60^{+0.05}_{-0.10}$\\[4pt]
   $L_{0.3-10} \cos \theta$ ($10^{38}$ erg s$^{-1}$)$^d$ 
      & $1.50^{+0.06}_{-0.06}$\\[4pt]
\enddata
\tablecomments{
$^a$: $N_{\rm {dbb}} = (r_{\rm{in}}/d_{10})^2 \cos \theta$, where $r_{\rm{in}}$ is the ``apparent'' inner disk radius in km, $d_{10}$ the distance to the galaxy in units of 10 kpc (here, $d_{10} = 300$), and $\theta$ is our viewing angle to the disk surface. The bolometric luminosity of the disk is $L \approx 4\pi r_{\rm {in}}^2 \, \sigma T_{\rm{in}}^4$ \citep{makishima86};\\
$^b$: the ``true'' inner-disk radius $R_{\rm{in}}$ is defined as $R_{\rm {in}} \approx 1.19\, r_{\rm in}$ for a standard disk \citep{kubota98}, and $r_{\rm {in}}$ was defined in Table note d;\\
$^c$: observed flux in the 0.3--8 keV band;\\
$^d$: $L_{0.3-10}$ is the un-absorbed luminosity in the 0.3--10 keV band, defined as $2\pi d^2/(\cos \theta)$ times the un-absorbed 0.3--10 keV model flux.\\
}
\label{tab:comb_result}
\end{deluxetable}

\begin{deluxetable}{lc}
\tablecaption{Best-fitting parameters of the {\it Chandra}/ACIS spectrum from ObsIDs 4743, fitted with the Cash statistics. The model is {\it tbabs}$_{\rm Gal}$ $\times$ {\it tbabs}$_{\rm int}$ $\times$ ({\it pow} $+$ {\it apec}). The Galactic column density was fixed at $N_{\rm {H,Gal}} = 4 \times 10^{20}$ cm$^{-2}$. Error ranges are 90\% confidence limits for one interesting parameter.}
\tablewidth{0pt}
\tablehead{
\colhead{Model Parameters} & \colhead{Values}
}
\startdata
   &   \\[-8pt]
$N_{\rm {H,int}}$   ($10^{20}$ cm$^{-2}$)   &  $ <4.5$  \\[4pt]
   $\Gamma$      &  $ 1.98^{+0.29}_{-0.29}$       \\[4pt] 
  $N_{\rm {po}}^a $ 
              &  $ 7.8^{+2.4}_{-2.1}$ \\[4pt]
   $kT_{\rm apec}$ (keV)     &  $ 1.20^{+0.34}_{-0.23}$\\[4pt] 
   $N_{\rm {apec}}^b$ &  $ \left(5.2^{+5.4}_{-3.4}\right) \times 10^{-6}$\\[4pt]
   C-stat/dof     &   $74.9/113$ (0.66)\\[4pt]
   $f_{0.3-8}$ ($10^{-14}$ erg cm$^{-2}$ s$^{-1}$)$^c$ 
       & $4.5^{+0.8}_{-0.7}$\\[4pt]
   $L_{0.3-10}$ ($10^{37}$ erg s$^{-1}$)$^d$ 
      & $5.7^{+1.0}_{-0.9}$\\[4pt]
\enddata
\tablecomments{
$^a$: units of $10^{-6}$ photons keV$^{-1}$ cm$^{-2}$ s$^{-1}$ at 1 keV.\\
$^b$: $N_{\rm {apec}} \equiv \left[10^{-14}/\left(4\pi d^2\right)\right]\,\int n_en_{\rm H} \, dV$, where $d$ is the luminosity distance in cm, and $n_e$ and $n_{\rm H}$ are the electron and H densities in cm$^{-3}$.\\
$^c$: observed flux in the 0.3--8 keV band\\
$^d$: un-absorbed 0.3--10 keV luminosity, defined as $4\pi d^2$ times the un-absorbed 0.3--10 keV model flux.\\
}
\label{tab:4743_result}
\end{deluxetable}

\subsection{Flux-resolved spectral modelling}
The next step of our spectral analysis is flux-resolved modelling. We want to determine whether and how the spectral shape changed between the higher-luminosity time bins and the eclipse ingress/egress time bins. Our first working hypothesis we shall test is that the accretion column (including both neutral and ionized gas) is higher immediately before and after the eclipses; we may also see a spectral change if different X-ray emission components come from different regions in the binary system, and some are more occulted or absorbed than others around an eclipse. For this objective, we include again ObsIDs 2030, 5197 and 22372, and leave out ObsID 4743, to avoid mixing different spectral states. 

In order to define a physically consistent count-rate threshold for the three epochs, we need to account for the decrease in ACIS-S sensitivity over the years. Based on the best-fitting {\it diskbb} model of the combined spectrum (Section 4.1), we estimate that 1 ct/ks in ObsID 2030 corresponds to 0.828 ct/ks in ObsID 5197 and 0.618 ct/ks in ObsID 22372. After binning the lightcurve to 500-s intervals, we defined various count-rate thresholds. For each threshold, we extracted and modelled two combined spectra from all the bins above and all those below that threshold. We report here the results for three choices of threshold: $10^{-2}$ ct s$^{-1}$ in ObsID 2030 (corresponding to $8.28 \times 10^{-3}$ ct s$^{-1}$ and $6.18 \times 10^{-3}$ ct s$^{-1}$ in the other two epochs); $2 \times 10^{-2}$ ct s$^{-1}$ ($1.66 \times 10^{-2}$ ct s$^{-1}$ and $1.24 \times 10^{-2}$ ct s$^{-1}$, respectively); $3 \times 10^{-2}$ ct s$^{-1}$ ($2.48 \times 10^{-2}$ ct s$^{-1}$ and $1.85 \times 10^{-2}$ ct s$^{-1}$, respectively). Other choices of thresholds lead to qualitatively similar results.

\begin{figure*}
\hspace{-0.5cm}
    \includegraphics[height=0.5\linewidth, angle=270]{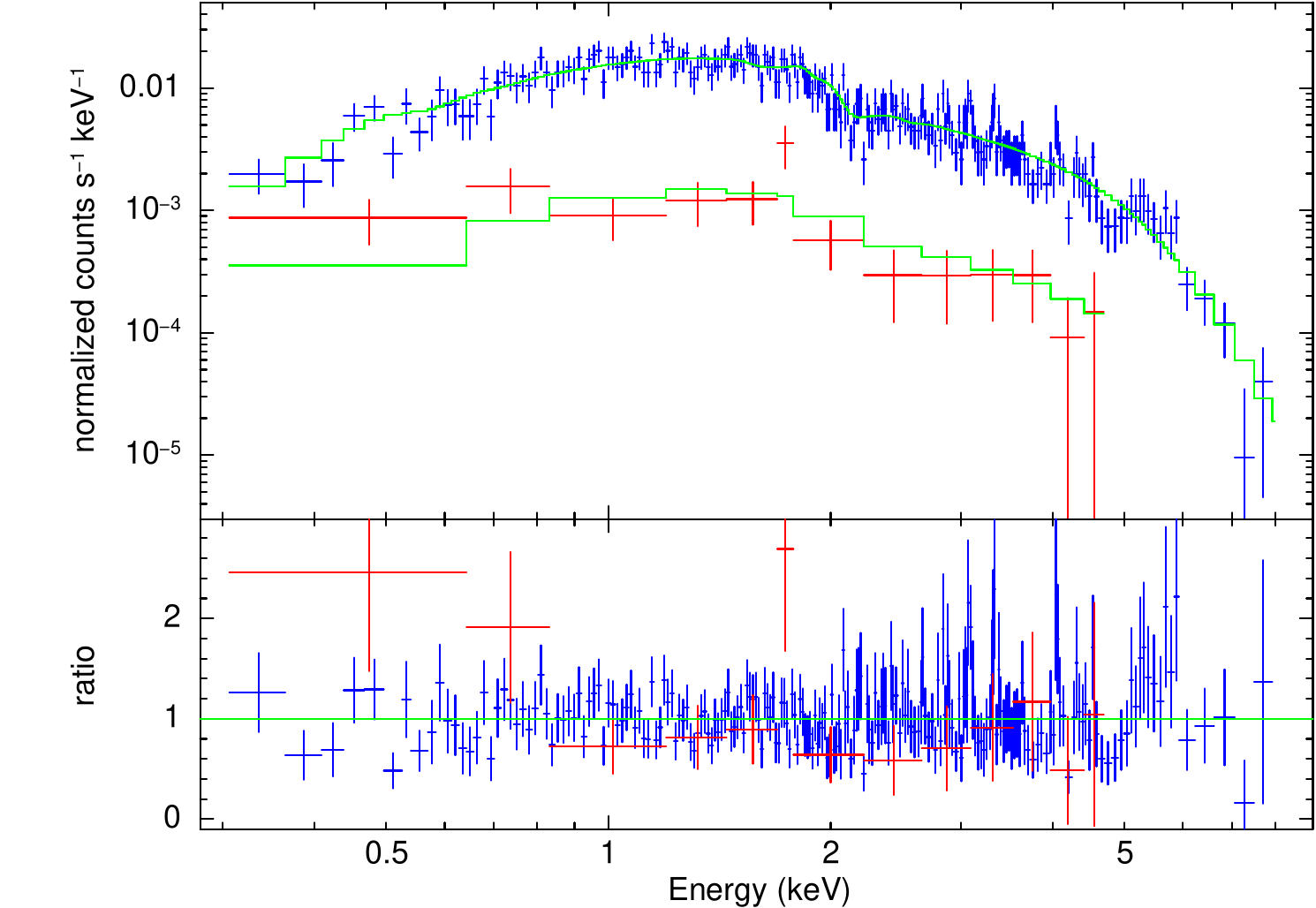}
    \includegraphics[height=0.5\linewidth, angle=270]{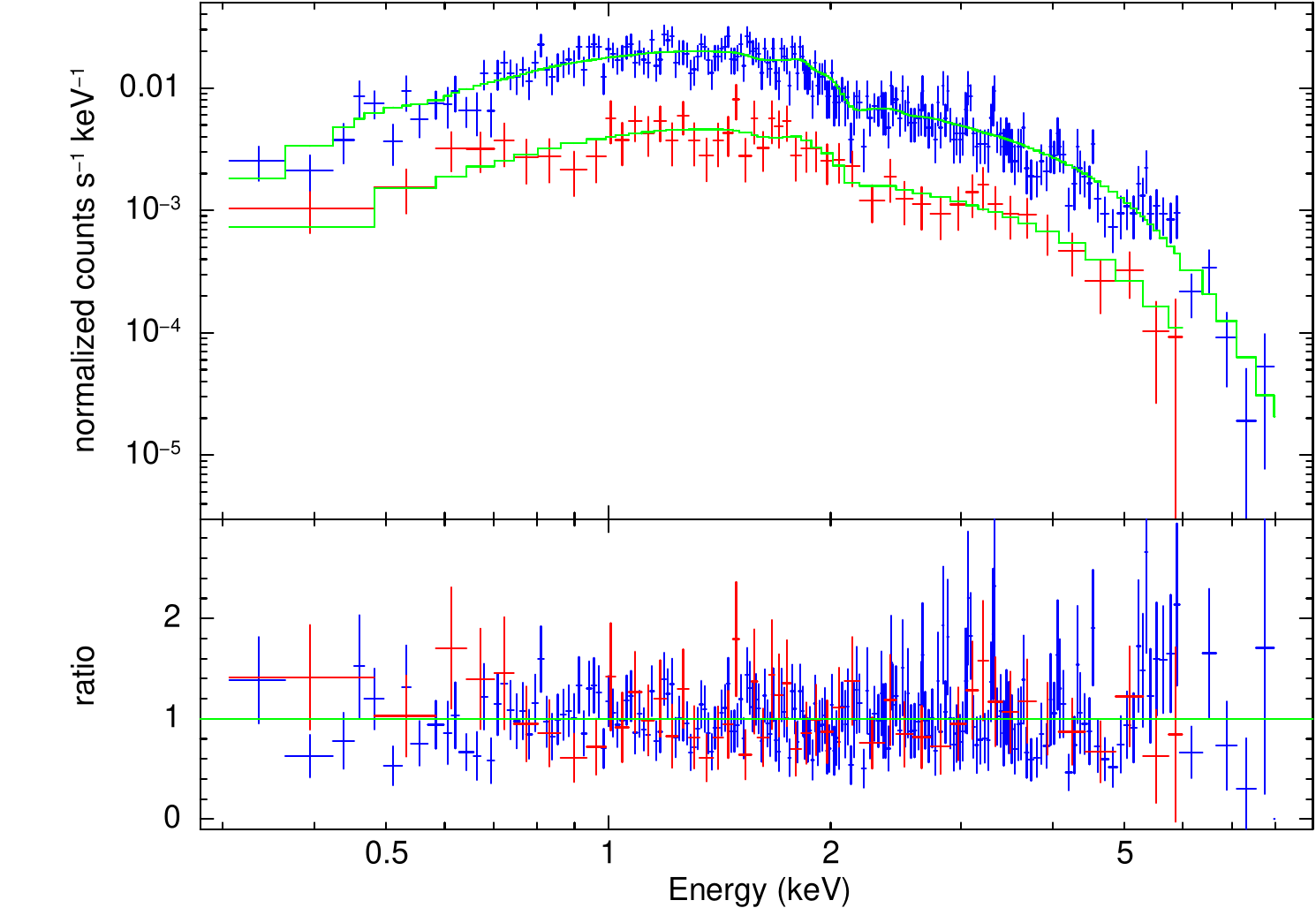}
    \caption{
\textcolor{black}{Left panel: {\it Chandra}/ACIS-S spectra above (blue datapoints) and below (red datapoints) a threshold corresponding to an equivalent count rate of 0.01 ct s$^{-1}$ in ObsID 2030. The spectra were simultaneously fitted with the Cash statistics; the datapoints have been rebinned to a signal-to-noise ratio $>$2.7 for plotting purposes. The fitted model is {\it tbabs}$_{\rm Gal}$ $\times$ {\it tbabs}$_{\rm int}$ $\times$ {\it pcfabs} $\times$ {\it diskbb}, with all parameters locked between the two spectra except for the covering fraction in {\it pcfabs}. The best-fitting covering fraction in the low-flux spectrum is $(91 \pm 2)$\% (C-stat $= 658.4/786$). A formally identical fit is obtained with the pure Compton scattering model {\it cabs}, which implies a column density of the scattering medium $N_{\rm H} = (3.0 \pm 0.3) \times 10^{24}$ cm$^{-2}$ in front of the low-flux spectrum.
Right panel: as in the left panel, for a threshold corresponding to an equivalent count rate of 0.03 ct s$^{-1}$ in ObsID 2030. The best-fitting partial-covering fraction is $(76 \pm 2)$\% for the low-flux spectrum. Alternatively, a Compton scattering model implies $N_{\rm H} = (1.8 \pm 0.1) \times 10^{24}$ cm$^{-2}$ (C-stat $= 825.7/874$ in both models).}  
}
\label{fig:cut_pcfabs}
\end{figure*}

The surprising result is that for any threshold, the difference between the higher and lower flux spectra is simply a constant scaling factor. There is no significant change of spectral shape, and no change in photoelectric absorption ($N_{\rm H}$ in the {\it tbabs} component), which would have removed preferentially soft photons. The two spectra of each pair are well modelled simultaneously with the same {\it diskbb} continuum, and a free partial-covering absorber ({\it tbabs}$_{\rm Gal}$ $\times$ {\it tbabs}$_{\rm int}$ $\times$ {\it pcfabs} $\times$ {\it diskbb} in {\sc xspec}). The column density in the {\it pcfabs} component is constrained to be $>$10$^{24}$ cm$^{-2}$ at the 90\% confidence level, which means a completely opaque absorber in the {\it Chandra} band. Thus, this is identical to simply placing a constant scaling factor in front of the {\it diskbb} component. 
\textcolor{black}{Specifically, for the 0.01 ct s$^{-1}$ threshold (Figure \ref{fig:cut_pcfabs}, left panel), the low-flux spectrum is absorbed by a Compton-thick medium with a covering fraction of $(91 \pm 2)$\%, while the covering fraction for the high-flux spectrum is assumed to be zero. The fit statistics is C-stat $= 658.4/786$. For the 0.02 ct s$^{-1}$ threshold, we obtain a covering fraction of $(84 \pm 2)$\% for the low-flux spectrum (combined fit statistics C-stat $= 776.8/87$). For the 0.03 ct s$^{-1}$ threshold (Figure \ref{fig:cut_pcfabs}, right panel), the covering fraction for the low-flux spectrum is $(76 \pm 2)$\% (C-stat $= 825.7/874$).}

A by-product of this exercise is that, by lifting the threshold level, we can estimate the average out-of-eclipse luminosity. We obtain an unabsorbed 0.3--10 keV luminosity $L_{\rm X} \cos \theta \approx 2.3 \times 10^{38}$ erg s$^{-1}$, which corresponds to $L_{\rm X} \approx 9 \times 10^{38}$ erg s$^{-1}$ at $\theta = 75^{\circ}$ and $L_{\rm X} \approx 1.3 \times 10^{39}$ erg s$^{-1}$ at $\theta = 80^{\circ}$.

It has been known since the early days of X-ray astronomy ({\it e.g.}, \citealt{mason85}) that an observed decrease in flux without substantial change in hardness and shape is usually caused either by a variable covering fraction in a clumpy medium ({\it i.e.}, an inhomogeneous medium with Compton thick clouds that cover only part of an extended X-ray emitting region), or by a highly ionized medium (no or few metal absorption lines and edges, with an approximately grey opacity dominated by Thomson scattering). Therefore, we also tried including a multiplicative ionized absorber component, {\it absori},  fitting the lower-flux and higher-flux spectral pairs with {\it tbabs}$_{\rm Gal}$ $\times$ {\it tbabs}$_{\rm int}$ $\times$ {\it absori} $\times$ {\it diskbb} models. Here, we allowed the column density of the {\it absori} component to vary between the two spectra of each pair, while all the other parameters were tied. We tried different ionization levels, up to $\xi = 5000$ (hard limit in {\sc xspec}). However, the resulting fits are significantly worse than in the partial covering model, with strong systematic residuals. For example, for the pair of spectra above and below the equivalent ObsID-2030 count rate of 0.03 ct s$^{-1}$, the best-fitting {\it absori} model at $\xi = 5000$ has a C-statistic of 900.5 for 876 degrees of freedom, compared with 825.7/874 for the partial covering model. Extrapolating the {\it absori} model beyond the {\sc xspec} limit, we estimate that we need an ionization parameter $\xi \gtrsim 10^4$ to reproduce the featureless decrease in flux seen in the data.

Finally, we replaced {\it absori} with the optically-thin Compton scattering model {\it cabs}, which accounts for the scattering of a fraction of X-ray photons out of the beam (model: {\it tbabs}$_{\rm Gal}$ $\times$ {\it tbabs}$_{\rm int}$ $\times$ {\it cabs} $\times$ {\it diskbb}). {\it cabs} attenuates the observed flux by a factor $\exp{\left[-N_{\rm H} \sigma_{\rm T}(E)\right]}$, where $\sigma_{\rm T}(E)$ is the Thomson cross section with Klein-Nishina corrections (essentially a constant over the {\it Chandra}/ACIS band). Fitting the spectrum with a {\it cabs} attenuation factor is formally equivalent to fitting a partial-covering fraction of a Compton-thick absorber, or simply multiplying the model by a free constant. As expected, we obtain statistically identical fits with {\it cabs} and {\it pcfabs} (Figures \ref{fig:cut_pcfabs}). 
\textcolor{black}{Specifically, for the 0.01 ct s$^{-1}$ threshold, the column density of the scattering medium is $N_{\rm H} = (3.0 \pm 0.3) \times 10^{24}$ cm$^{-2}$ for the low-flux spectrum, while $N_{\rm H} \equiv 0$ for the high-flux spectrum (C-stat $= 658.4/786$). For the 0.02 ct s$^{-1}$ threshold, the scattering $N_{\rm H} = (2.3 \pm 0.2) \times 10^{24}$ cm$^{-2}$ (C-stat $= 776.8/875$). For the 0.03 ct s$^{-1}$ threshold, $N_{\rm H} = (1.8 \pm 0.1) \times 10^{24}$ cm$^{-2}$ (C-stat $= 825.7/874$).}

\section{Optical results}\label{sec:hst}
The point-like optical counterpart stands out as one of the brightest sources within a few 10s of pc of the X-1 position (Figure \ref{fig2}), both in the near-UV and in the near-IR, but is relatively faint in the $V$ band (Table \ref{tab:HST}). An approximate conversion to standard colors indicates $U-B \approx -1.2$ mag, and $V-I \approx 1.6$ mag. There are no stars with this type of bimodal colors, which are inconsistent with any single thermal model. 

\begin{figure}
\hspace{-0.5cm}
    \includegraphics[height=1.05\linewidth, angle=270]{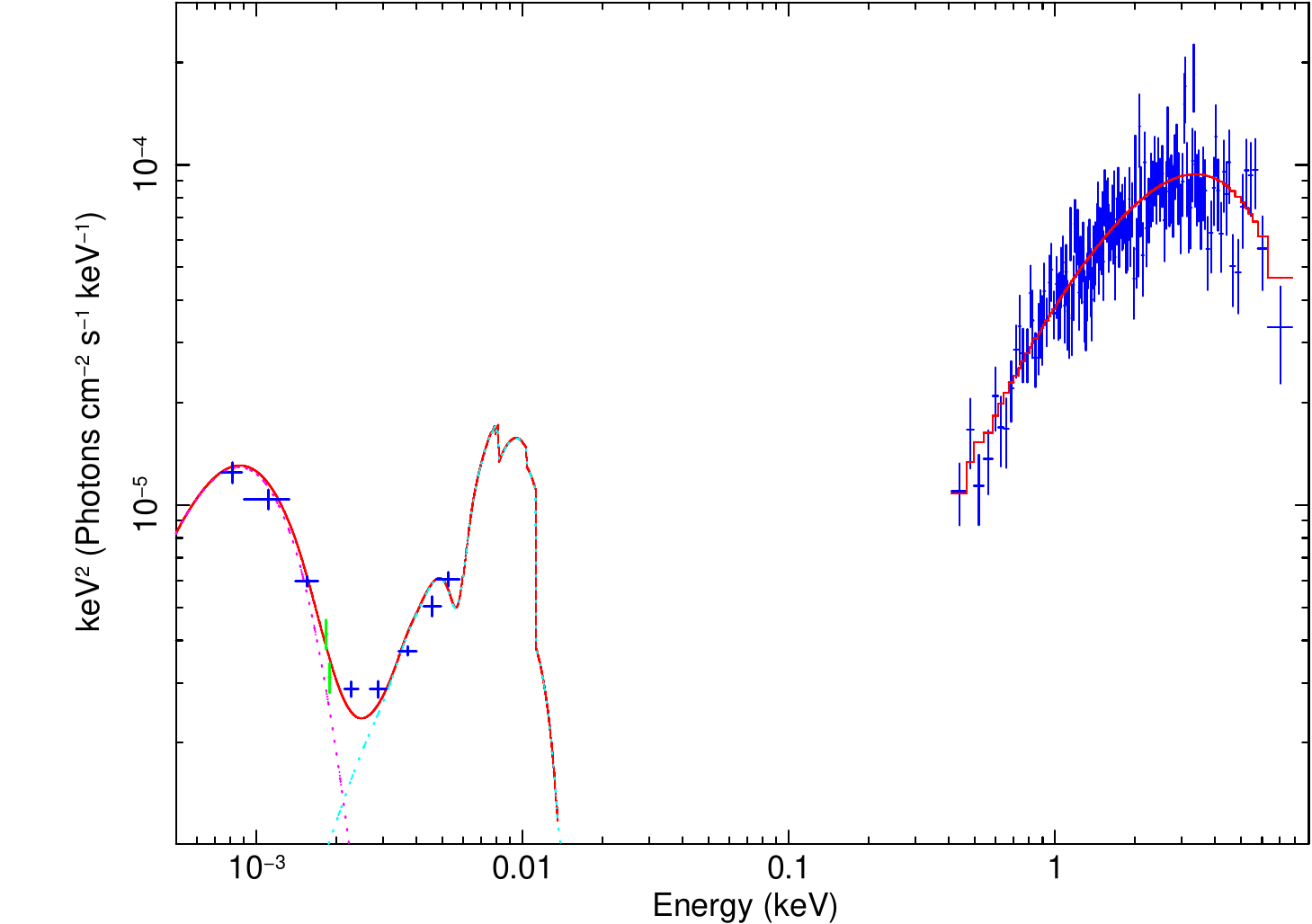}
    \caption{SED plot showing that the optical data require two bb components: one at a characteristic temperature of $\approx (90,000 \pm 30,000)$ K and radius $\approx (2.1 \pm 0.3) R_{\odot}$ (strongly dependent on the poorly constrained reddening); the other at $T \approx (2700 \pm 300)$ K and $R \approx (420 \pm 50) R_{\odot}$.  Blue datapoints are from {\it Chandra} and from broad-band and medium-band filters of {\it HST}/WFC3 (Table \ref{tab:HST}). In addition, we have plotted (green datapoints) the flux density measured in the two narrow-band {\it HST}/WFC3 filters with the highest signal to noise ratio (F657N and F673N).}
    \label{fig:full_sed}
\end{figure}

The first hypothesis we tested was that the blue/UV emission may come from the irradiated accretion disk. We re-fitted the combined X-ray spectrum from ObsIDs 2030, 5197 and 22372 with a {\it diskir} model in {\sc xspec}. In the 0.3--8 keV band, {\it diskir} is consistent with the fitting results of {\it diskbb} discussed earlier ({\it e.g.}, $kT_{\rm in} \approx 1.3$ keV and $R_{\rm in} \cos \theta \approx 25$ km); however, {\it diskir} includes the calculation of the outer disk emission down to the optical/IR band. We found that the observed blue/UV luminosity is $\sim$10\% of the soft X-ray luminosity (Figure \ref{fig:full_sed}). This fraction is too high to originate from disk reprocessing of X-ray photons according to the {\it diskir} model. It would require the disk to intercept $\sim$20\% of the X-ray photons (including the albedo term), a level that is not self-consistent with a standard disk model. For plausible reprocessing fractions $\la$0.01, the model irradiated disk contribution to the observed blue/UV emission is $<$10\%. The disk contribution to the red/IR emission is much lower, if we assume an outer truncation radius $R_{\rm out} \sim 10^4 r_{\rm in}$ (a relatively small outer radius is justified by the short binary period, see Section 6.1). 

Next, we modelled the near-UV/optical/near-IR emission with a double thermal model ({\it bbodyrad} $+$ {\it bbodyrad}), in addition to the (small) {\it diskir} contribution \textcolor{black}{in {\sc xspec}}. This model includes several free parameters: temperature and radii of the two blackbody components, outer disk radius, reprocessing fraction, intrinsic reddening (which may or may not be the same for the blue and red components). We only have eight broad- or medium-band {\it HST}/WFC3 datapoints. Thus, there will inevitably be a degeneracy of acceptable solutions (in particular, reddening and blackbody temperature of the hotter emitter are strongly degenerate). To reduce such degeneracy, we froze the outer disk radius at $R_{\rm out} = 10^4 r_{\rm in}$ and the reprocessing fraction at $f = 5 \times 10^{-3}$, values that imply a negligible disk contribution to the optical emission.

Our objective here is simply to give an order of magnitude estimate of the size and temperature of the colder and hotter emitter. For low values of the total (intrinsic plus Galactic line-of-sight) reddening $E(B-V) \approx 0.10$ mag (consistent with the X-ray spectral models), we obtained good fits for a blue emitter with a
temperature $T_{\rm hot} \approx 60,000$--80,000 K and a radius $R_{\rm hot} \approx (2.0 \pm 0.5) R_{\odot}$. For $E(B-V) \approx 0.25$ mag, the preferred temperature range is $T_{\rm hot} \approx 100,000$--120,000 K while the radius is unchanged. Bolometric luminosities of the hotter blackbody component are in the range $\log \left(L_{\rm bol}/L_{\odot}\right) \approx 5.2$--6.0, depending on the choice of reddening.
The cooler emitter has acceptable solutions for $T_{\rm cold} \approx 2500$--3000 K, radius $R_{\rm cold} \approx 400$--500 $R_{\odot}$ and bolometric luminosity $\log \left(L_{\rm bol}/L_{\odot}\right) \approx 4.0$.

At a distance modulus of $\approx$27.4 mag, the dereddened absolute brightness of the blue component is $M_{\rm F438W} \approx M_B \approx -3.9$ mag to $-3.3$ mag for a range of plausible extinctions. In the $V$ band, the observed flux receives comparable contributions from both the hotter and the cooler emitter (Figure \ref{fig:full_sed}). Assuming for example an exact half/half split between the two components, the dereddened $V$-band absolute brightness is $M_{\rm F547M} \approx M_V \approx -3.3$ mag to $-2.8$ mag for each of the two components. At longer wavelengths, we estimate a dereddened $I$-band absolute brightness $M_{\rm F814W} \approx M_I \approx -5.3$ mag to $-5.0$ mag. 

Finally, we measured the brightness in the two narrow-band UVIS filters F657N and F673N. We plotted their flux densities in Figure \ref{fig:full_sed}, because they help filling a data gap in the $R$ band, but did not use those two datapoints for our spectral fitting. We do not find evidence for significant brightness at H$\alpha$ and [\ion{S}{2}] $\lambda 6716,6731$ in excess of the modelled continuum. 

\section{Discussion}\label{sec:discussion}

\subsection{Orbital parameters from eclipse duration}

For circular orbits, the eclipse fraction $\Delta_{\rm ecl}$ is related to the donor star radius $R_2$, the binary separation $a$ and the \textcolor{black}{orbital plane inclination} $i$ by the relation
\begin{equation}
    \sin i \cos \left(\pi \Delta_{\rm ecl}\right) = \left[1 - \left(R_2/a\right)^2\right]^{1/2}
\end{equation}
\citep{porquet05,chanan76}.
Given the high and persistent (over at least two decades) X-ray luminosity, and the short binary period, we can plausibly assume that the donor star is close to filling its Roche lobe. Thus, applying \cite{eggleton83}'s approximation for $R_2/a$, we obtain numerical solutions for the mass ratio $q = M_2/M_1$ as a function of $i$ and $\Delta_{\rm ecl}$:
\begin{equation}
    \frac{\ln \left(1+q^{1/3}\right)}{q^{2/3}} = 
    \frac{0.49}{\sqrt{1-\sin^2 i \cos^2 \left(\pi \Delta_{\rm ecl}\right)}} - 0.60
\end{equation}
\citep{porquet05}.
The minimum value of $q$ corresponds to $i = 90^{\circ}$. For the value of $\Delta_{ecl}$ determined for X-1 (Section 3.2), we have $q_{\rm min} = 3.0^{+1.7}_{-1.1}$. The importance of this result is that it rules out a low-mass X-ray binary solution for X-1: the donor star must be more massive than the accretor. For a typical NS accretor, the best-fitting estimate suggests $M_2 \ga 4 M_{\odot}$, and for a stellar-mass BH accretor, $M_2 \ga 15 M_{\odot}$.

From the period-density relation 
\begin{equation} 
    P\rho_2^{1/2} = 0.1375\left(\frac{q}{1+q}\right)^{1/2}\left(R_2/a\right)^{-3/2}
\end{equation}
\citep{eggleton83}, where $P$ is in days and $\rho_2$ is g cm$^{-3}$, a mass ratio $q \approx 3$ corresponds to $\rho_2 \approx 5.9$ g cm$^{-3}$. This further rules out main-sequence or giant/supergiant stars, and confirms that the donor star is a stripped star.

\subsection{Mass and spectral type of the donor star}
We shall assume here for simplicity that $i \approx 90^{\circ}$, and discuss alternative scenarios consistent with $q\approx3$ inferred from the eclipse duration (Section 6.1), and with the optical brightness derived from the {\it HST} images (Section 5). As illustrative examples, for a NS scenario, we take $M_1 + M_2 \approx 1.5 + 4.5 = 6 M_{\odot}$, and for a BH scenario, we assume $M_1 + M_2 \approx 7 + 21 = 28 M_{\odot}$. For the same best-fitting binary period, simple Newtonian physics suggests a binary separation $a \approx 1.5 \times 10^{11}$ cm in this NS scenario, with $R_2 \approx 1.0 R_{\odot}$ \citep{eggleton83}; the BH scenario implies $a \approx 2.5 \times 10^{11}$ cm, $R_2 \approx 1.7 R_{\odot}$.

Are such values consistent with stellar evolution on the one hand, and with the observed optical brightness on the other? Evolutionary tracks of intermediate-mass stars at Large-Magellanic-Cloud metallicity ($Z = 0.006$), 
stripped of their hydrogen envelope through binary interactions, show \citep{gotberg18} that stars with initial masses $M_{\rm i} \approx 18 M_{\odot}$ and stripped mass $M \approx 7 M_{\odot}$ have effective radius $R_{\rm eff} \approx 1.3 R_{\odot}$, effective temperature $T_{\rm eff} \approx 85,000$ K, absolute brightness $M_{B} \approx -2.1$ mag, $M_{V} \approx -1.6$ mag, and bolometric luminosity $\log \left(L_{\rm bol}/L_{\odot}\right) \approx 4.9$. This is still at least 1.2 mag fainter in $B$ and $V$ than the observed blue component, and at least a factor of 3 smaller in $L_{\rm bol}$. \cite{gotberg18} do not tabulate evolutionary tracks for more massive stripped stars, but, by extrapolating from the trend of their highest-mass tracks, we estimate that stars with $M \ga 10 M_{\odot}$ ($M_{\rm i} \ga 25 M_{\odot}$) will be consistent with our observed optical brightness and modelled $L_{\rm bol}$. 

He stars more massive than $\sim$10 $M_{\odot}$ are generally classified as classical WR stars. Evolutionary tracks \citep{hamann04,todt15,grafener17} as well as observations \citep{hainich14,hainich15,hamann06} show that only the WN2--WN3 sub-types of WR stars are consistent with the relatively faint $M_{V} \sim -3$ mag inferred for the blue component of the X-1 counterpart. Such early-type nitrogen-sequence WR stars have current masses $\sim$12--16 $M_{\odot}$, bolometric luminosity $\log \left(L_{\rm bol}/L_{\odot}\right) \approx 5.3$--5.6, and $T_{\rm eff} \sim 100,000$--120,000 K. 

In summary, comparing the observed optical brightness (with the caveat that a simple blackbody model is only a rough approximation to real stellar spectra) with the predictions from evolutionary tracks, we argue that the blue emitter is most consistent with a He star/WR star with a current mass $10 \la \left(M/M_{\odot}\right) \la 16$. Masses near the upper limit of this mass range are required if the accretor is a BH, to explain the long duration of the X-ray eclipse. For a NS accretor, the eclipse duration is consistent with the whole mass range.

Finally, we investigate the origin of the red component (Section 5). At $M_{I} \approx -5$ mag, it is one magnitude brighter than the tip of the red giant branch, but 1--2 magnitudes fainter than typical red supergiants. Instead, it is perfectly consistent with the asymptotic giant branch (AGB) of stars with initial masses $\sim$2--3 $M_{\odot}$, at an age of $\sim$200--1,000 Myr (Padova isochrones\footnote{http://stev.oapd.inaf.it/cgi-bin/cmd}: \citealt{bressan12,marigo13,marigo17,nguyen22}; see also \citealt{cinquegrana22}). Its temperature $T \approx 2600$--3000 K is consistent with an AGB star but too cold for a red supergiant ($T \sim 3500$--4000 K). Moreover, AGB tracks (especially in their thermally pulsing phases) predict bolometric luminosities $\log \left(L_{\rm bol}/L_{\odot}\right) \approx 4.0$, similar to what we estimated from our blackbody fit.

The presence in the same system of a young He star/WR star (age $\la$20 Myr) and an intermediate-mass AGB star (age $\ga$200 Myr) is baffling. They could not have been formed together. The short binary period required by the X-ray light-curve implies that the 500-$R_{\odot}$ AGB star cannot be the mass donor. Short-period X-ray binaries with persistent $L_{\rm X} \sim L_{\rm Edd}$ are rare and short-lived systems in the local universe (expected lifetime $\la$ few Myr), and thermally pulsing AGB stars are also rare and short-lived (a few Myr). The chance of having an unrelated AGB star spatially aligned (to within $0\farcs{1}$) with the true donor star of X-1 appears low. Thus, it is worth considering alternative scenarios for the two-component optical emission. For example, WR stars produce dust and sometimes show double peaked spectral energy distributions, with a blue and red component \citep{lau19}. However, dust sublimates at a temperature of $\approx$1300--2000 K, depending on chemical composition \citep{phinney89,kobayashi11}; this is significantly colder than the red component observed in the {\it HST} images. A less implausible scenario is that the red component is a circumbinary (CB) disk or ring, perhaps the remnants of a recently ejected common envelope between the WR star and compact accretor. This scenario is viable if the large CB structure intercepts and re-emits a few per cent of the combined X-ray luminosity from the inner accretion disk and UV luminosity from the WR donor star. 

The possibility of a leftover CB disk after common envelope ejection has previously been examined for the case of a white dwarf in the envelope of an AGB star \citep{kashi11,demarco11,passy11,reichardt19}. In general, for binary systems inside CB disks, tidal transfer of angular momentum from the binary to the disk contributes to a further decrease of the binary separation \citep{shi12,ma09,chen06,artymowicz94,artymowicz91}, as well as expansion of the CB disk radius \citep{kashi11,chen17}. Most of the theoretical work on CB disks has explored the scenario where the inner disk has a net transfer of mass onto the inner binary, in particular onto the less massive of the two components \citep[{\it e.g.},][]{bowen18,farris15,gold14,dorazio13, shi12,artymowicz96}. In the case of NGC\,4214 X-1, given the expected strong wind mass losses from the WR star, it is more likely that we are in the opposite scenario, of a CB disk being fed (positive mass transfer from the binary to the CB disk) at its inner radius by the binary donor star, and expanding outwards \citep{dubus02}. A substantial problem with the CB scenario is how to justify a characteristic size of several 100 $R_{\odot}$ for the optically thick emitter, compared with a binary separation of $\approx$2 $R_{\odot}$. 
A modelling study of the dynamical stability, accretion rate and size and optical thickness of post-common-envelope CB disks is beyond the scope of this work.

\subsection{Reasons for the extended eclipse egress}
A distinctive feature of the X-ray eclipses in NGC\,4214 X-1, visible in individual cycles (Figure \ref{fig:eclipse_in.png}) and, more clearly, in the folded light-curve (Figure \ref{fig:phase_count_6p}), is the longer duration of egresses compared with ingresses. It takes $\la$1000 s for the flux to drop from peak to minimum value, and, on average, $\approx$4000 s for the flux to recover gradually to peak value after the eclipse. The eclipse profile changes from cycle to cycle, which suggests that it is due to transient and stochastic properties of the medium between the two stars. An egress slower than the ingress is common in the small group of candidate short-period WR systems: Circinus Galaxy X-1 \citep{qiu19b}, IC\,10 X-1 \citep{laycock15}, NGC\,300 X-1 \citep{binder11,binder15}, NGC\,4490 X-1 \citep{esposito13}, Cygnus X-3 \citep{zdziarski12,antokhin22}. Of those sources, the three systems with the most dramatic eclipse asymmetry and most extended egress duration are also those with the highest luminosity, $L_{\rm X} \gtrsim 10^{39}$ erg s$^{-1}$: Circinus Galaxy X-1, NGC\,4490 X-1 and NGC\,4214 X-1. This asymmetry is not common, instead, in ordinary wind-accreting HMXBs with supergiant donors \citep{Falanga15}. 

We cannot explain the slow egress simply as a gradual decrease of the column density of a cold photo-electric absorber, because of the lack of spectral evolution during egress (Section 4.2). The same lack of spectral evolution is also a distinctive feature of Circinus Galaxy X-1 \citep{qiu19b}. Our spectral modelling is consistent with either a variable partial covering by opaque clouds, or variable column density of a completely ionized absorber (grey opacity), which may scatter part of the X-ray photons out of our line-of-sight during egress (but not during ingress). The ionization parameter required to explain the observed spectra ($\xi \gtrsim 10^4$, Section 4.2) is easily reached in this compact, luminous system, because $\xi \approx 10^5 L_{39} \, n^{-1}_{12} \, r^{-2}_{11}$ (where $L_{39} \equiv L_{\rm X}/10^{39}$ erg s$^{-1}$, $n_{12} \equiv n/10^{12}$ cm$^{-3}$, $r_{11} \equiv r/10^{11}$ cm).

The difficulty of the partial covering scenario is that the occulting clouds must be much smaller than the X-ray emitting region (the innermost disk region, within $\lesssim$ a few 100 km), otherwise the X-ray source would only be either fully covered or fully visible. 
There is no empirical evidence so far for the existence of clumps or winds of that small size. Alternatively, larger, fast-moving clumps may indeed cause the X-ray source to flicker only between fully visible and fully occulted intervals during egress, but on timescales shorter than our time resolution (set by the low observed count rate). If the fraction of fully occulted time within a time resolution element decreases, the process appears to us as a ``gradual'' flux recovery. This scenario would be empirically indistinguishable from the partial covering scenario.

The reason for the apparently grey opacity and the reason for the asymmetric distribution of the absorber in ingress and egress are likely to be physically connected. The observed asymmetry implies that the additional absorber must be either leading the compact object or trailing the donor star. The former scenario could be explained as a bow shock, caused by the supersonic orbital motion of the compact object inside the WR wind \citep{hunt71,theuns93,wang14}. This is the scenario invoked by \cite{qiu19b} for Circinus Galaxy X-1 and by \cite{antokhin22} for Cyg X-3. The latter scenario relies on the presence of a \textcolor{black}{``shadow wind'' in the system (\citealt{blondin94} and \citealt{blondin95} for hydrodynamical simulations; \citealt{laycock15b} for an application of this scenario to the WR system IC\,10 X-1).} In a compact, luminous system, the X-ray photons from the compact object suppress or strongly decrease the radiatively driven wind from the donor star, except from its shaded side \citep{blondin94,vilhu21}. Coriolis forces deflect the shadow wind towards the trailing side of the donor star: this is the reason why it appears between us and the compact object in eclipse egress but not in ingress. As the wind leaves the WR shadow, it stalls and gets highly ionized \citep{blondin94}. In our grey opacity scenario, this is the ionized medium responsible for scattering some of the X-ray photons out of our line of sight. Further analysis of the integrated column density of the shadow wind is beyond the scope of this work.

\section{Conclusions}
We combined new and archival {\it Chandra} and {\it HST} data for a study of the short-period, eclipsing X-ray binary NGC\,4214 X-1. We confirmed that the source is still active and still showing eclipses, with an out-of-eclipse luminosity $L_{\rm X} \approx 10^{39}$ erg s$^{-1}$, 19 years after the first {\it Chandra} observation. \textcolor{black}{Persistently active sources at such high luminosities tend to have high-mass donors, although here we do not have direct information (such as radial velocity curves) to constrain the mass of the donor star.}
We refined the estimate of the orbital period, $P \approx (12940 \pm 270)$ s, and the average eclipse duration, $\Delta T_{\rm ecl} \approx (2050 \pm 175)$ s. From the high eclipse fraction, $\Delta_{\rm ecl} \approx 0.16$, we derive a minimum mass ratio $q_{\rm min} > 2$ (and, most likely, $q \gtrsim 3$), which again suggests that the system is an HXMB. At the same time, the short binary period rules out main sequence or supergiant donors, and is only consistent with a WR star or an intermediate-mass stripped He star.

From the {\it HST}/WFC3 observations, we discovered the likely optical counterpart, at an apparent brightness $V \approx 24$ mag and an absolute magnitude $M_V \approx -3$ mag, $M_I \approx -5$ mag. The optical source is inconsistent with a single thermal component, and consists instead of two clearly distinct components: a blue one, with $T_{\rm hot} \gtrsim 60,000$ K and characteristic radius $R_{\rm hot} \approx 2 R_{\odot}$, and a red one, with $T_{\rm cold} \approx 2,700$ K and characteristic radius $R_{\rm hot} \approx 400 R_{\odot}$. The blue emission is consistent with a WR star, in particular from the compact, early-type WN class, with stripped mass $\sim$12--16 $M_{\odot}$.  The red component is either an unrelated AGB star from an older stellar population, projected on the sky at the same location as the true optical counterpart (within $0\farcs{1}$), or it is emission from a CB disk, perhaps fed by the WR wind.

Our X-ray timing and spectral study highlights two properties that seem to characterize the rare class of luminous (near- or super-Eddington) WR HMXBs: i) a highly asymmetric eclipse profile, with a sharp ingress and a longer egress; ii) a lack of spectral evolution during the slow egress, as the observed flux increases. The asymmetry suggests that most of the absorbing material is either leading the compact object, or trailing the donor star, along their orbital motion. We speculate that, in the first scenario, the absorbing material could be associated with a bow shock in front of the compact object; in the latter scenario, it could be the shadow wind from the un-irradiated face of the donor star, which is launched in the radial direction but trails the star because of its orbital motion. The strong X-ray luminosity and small binary separation ensure that as soon as this wind component leaves the shadow region, it gets completely ionized ($\xi \gtrsim 10,000$). This would explain why its effect on the observed X-ray spectrum is via Thomson scattering of photons out of our line of sight (grey opacity) rather than photo-electric absorption.

We cannot rule out either a NS or a BH nature of the compact object. However, a near-Eddington BH interpretation appears more consistent with the observed X-ray spectrum, which is well modelled by a disk-blackbody with $kT_{\rm in} \approx 1.3$ keV and $R_{\rm {in}}\sqrt{\cos \theta} \approx 25$ km. A BH nature of the compact object would also be consistent with the suggestion that BHs are more likely to survive a common envelope phase, prelude to the formation of short-period WR X-ray binaries \citep{vandenheuvel17}. In any case, NGC\,4214 X-1 will eventually produce a double compact object binaries, tight enough to merge via gravitational wave emission on timescales of a few $10^7$ yr. Only a handful of such systems have been detected so far, partly because of observation biases: that is, we may not recognize similar systems (bright HMXBs) when they are seen more face-on, without observable X-ray eclipses.

\begin{acknowledgments}

This research was supported by the National Science Foundation of
China (NSFC) under grant numbers 11988101, 11933004, 12073029; by the National Key Research and Development Program of China (NKRDPC) under grant numbers 2019YFA0405504 and 2019YFA0405000; and by the Strategic Priority Program of the Chinese Academy of Sciences under grant number XDB41000000. Support for this work was also provided by the National Aeronautics and Space Administration through Chandra Award Number
 GO0-21029X issued by the Chandra X-ray Center, which is operated by the Smithsonian Astrophysical Observatory for and on behalf of the National Aeronautics Space Administration under contract NAS8-03060.

We thank Ryan Lau, Jifeng Liu, Manfred Pakull, Yanli Qiu, Varsha Ramachandran and Andreas Sander for their comments and suggestions. We also acknowledge fruitful group discussions on Wolf-Rayet stars at the International Space Science Institute meeting in Bern, April 2023 (team 512, ``Multiwavelength View on Massive Stars in the Era of Multimessenger Astronomy'', led by Lidia Oskinova). RS also thanks Lei Zhang for VPN support during his stay in China, which made this study possible. 

The scientific results reported in this article are based to a significant degree on observations made with the {\it Chandra X-ray Observatory}. 
This research is also based on observations made with the NASA/ESA {\it Hubble Space Telescope} obtained from the Space Telescope Science Institute, which is operated by the Association of Universities for Research in Astronomy, Inc., under NASA contract NAS 5–26555. These observations are associated with Program IDs 11360 (PI: R.~O'Connell) and 17225 (PI: D.~Calzetti). All the {\it Chandra} and {\it Hubble} data used in this study are downloadable from their respective public archives.

We made use of the online interface of the Potsdam Wolf-Rayet Models (PoWR), at http://www.astro.physik.uni-potsdam.de/PoWR.html, and the online interface of the PAdova TRieste Stellar Evolutionary Code (PARSEC) V2.0, at http://stev.oapd.inaf.it/cgi-bin/cmd.

\end{acknowledgments}

\bibliography{main}{}
\bibliographystyle{aasjournal}

\appendix

\section{X-ray properties of SNRs in NGC\,4214}
As a serendipitous result, we searched for X-ray counterparts of the seven SNRs identified by \cite{dopita10} in this galaxy. Six of them (all except the 4th one in their list) are recovered in the combined {\it Chandra} image (Figure \ref{fig1}), which has a total exposure of 141 ks. We assumed that the luminosity of the SNRs remained constant across the four {\it Chandra} observations, and that their spectra could be approximated with a thermal plasma model ({\it apec} model in {\sc xspec}) at $kT \approx 0.5$ keV. We fitted the spectra of two SNRs (SNR2 and SNR5) with high signal-to-noise ratio and relatively low contamination from diffuse hot gas emission, and provided a rough estimate of the luminosities of the other four SNRs based on their scaled count rates (Table \ref{tab:SNR}).
\textcolor{black}{We find five SNRs with luminosities $>$10$^{36}$ erg s$^{-1}$, up to a maximum luminosity $\approx 3 \times 10^{36}$ erg s$^{-1}$. Taking into account the small number statistics, this is perfectly consistent with the expected number and luminosities of such sources, by comparison with the high-end of the X-ray SNR luminosity functions seen in the LMC \citep{maggi16} and in M\,33 \citep{garofali17}, which have star formation rates $\approx$3--4 times higher than NGC\,4214 \citep{harris09,antoniou16,williams13}.}

\restartappendixnumbering

\begin{deluxetable}{lccccc}
\tablecaption{X-ray properties of the optically identified SNRs \citep{dopita10} recovered in the {\it Chandra} observations of NGC\,4214. \label{tab:SNR}}
\tablewidth{0pt}
\tablehead{
\colhead{Source} & \colhead{RA} & \colhead{DEC} & \colhead{\textcolor{black}{Net Rate}} & \colhead{0.3--7 keV Flux} & \colhead{0.3--10 keV Luminosity} \\[-5pt]
 & (hms) & (hms) & ($10^{-5}$ ct/s) & ($10^{-15}$ erg cm$^{-2}$ s$^{-1}$) & ($10^{36}$ erg s$^{-1}$) }
\startdata
SNR1 & 12:15:38.963 & +36:18:58.85 & $9.0\pm3.1$ & $0.5 \pm 0.2$ & $0.8 \pm 0.3$\\
SNR2 & 12:15:40.000 & +36:18:40.72 & $36.5\pm5.6$ & $2.1\pm 0.3$ & $3.1 \pm 0.4$\\
SNR3 & 12:15:40.008 & +36:19:36.00 & $34.7\pm7.0$ & $2.1\pm 0.4$ & $3.1 \pm 0.5$\\
SNR5 & 12:15:41.136 & +36:19:15.04 & $21.1\pm4.4$ & $1.1 \pm 0.2$ & $1.4 \pm 0.3$\\
SNR6 & 12:15:42.456 & +36:19:47.78 & $14.5\pm4.0$ & $0.8 \pm 0.2$ & $1.2 \pm 0.3$\\
SNR7 & 12:15:45.639 & +36:19:41.50 & $19.6\pm4.4$ & $1.0 \pm 0.2$ & $1.3 \pm 0.3$\\
\enddata
\end{deluxetable}

\end{document}